\documentclass[acp,manuscript]{copernicus}

\renewcommand{\thesubsection}{\thesection.\alph{subsection}}

\usepackage{latexsym,euscript,textcomp}
\usepackage{color,graphics,epsf,dcolumn}
\usepackage{epstopdf,ulem}
\usepackage{blindtext}
\usepackage{threeparttable}
\usepackage{xcolor}

\newcommand{\beq}{\begin{equation}}
\newcommand{\eeq}{\end{equation}}
\newcommand{\pt}{\partial}

\begin{document}
\nolinenumbers

\title{Condensation mass sink and intensification of tropical storms}


\Author[1,2]{Anastassia M.}{Makarieva}
\Author[1]{Andrei V.}{Nefiodov}

\affil[1]{Theoretical Physics Division, Petersburg Nuclear Physics Institute, Gatchina  188300, St.~Petersburg, Russia}
\affil[2]{Institute for Advanced Study, Technical University of Munich,  Garching 85748, Germany}


\runningtitle{Condensation mass sink}

\runningauthor{A.M. Makarieva and A.V. Nefiodov}

\correspondence{A. M. Makarieva (ammakarieva@gmail.com)}

\received{}
\pubdiscuss{} 
\revised{}
\accepted{}
\published{}


\firstpage{1}

\maketitle

\begin{abstract}
{\large
Intensification of tropical storms measured as the central pressure tendency represents a subtle imbalance, of the order of $10^{-3}$, between the inflow and outflow of air in the storm core. Factors driving this imbalance, especially in cases of rapid intensification, remain elusive. Here, using an analysis of intensification rates and precipitation in North Atlantic cyclones, it is shown that the storms on average deepen at a rate with which maximum local precipitation removes  mass from the atmospheric column. Means for lifetime maximum intensification rate  and maximum concurrent precipitation (multiplied by the acceleration of gravity) are, respectively, $23$ and $17$ hPa~day$^{-1}$. This equivalence is not limited to average values: both intensification rates and precipitation have the same dependence on the inverse radius of maximum wind. It is further shown using a numerical model that with the mass sink switched off, storms driven by sensible and latent heat alone either do not develop at all or develop significantly more slowly reaching lower maximum intensities. It is discussed that the conclusions of previous studies about the relative insignificance of the mass sink arose from a long-standing  misinterpretation of mass nonconservation assessments for assesments of the actual impact of the mass sink on storm dynamics. Condensation mass sink provides for a fundamental positive feedback between surface pressure and vertical velocity that  was earlier shown to be instrumental in analytical descriptions of storm intensification. This feedback allows the storm circulation to get more compact during intensification in contrast to modeled heat-driven storms that increase their radius of maximum wind as they intensify. These findings indicate that the condensation mass sink is a dominant process governing the dynamics of tropical storms. }
\end{abstract}

{\large
\section{\large Introduction}
\label{intr}

Rapid intensification of tropical storms defies reliable forecasting, threatens human lives and property, and largely remains enigmatic. \citet{holliday79} analyzed maximum deepening rates (the largest daily drop in central surface pressure during the storm{\textquoteright}s lifetime) for $305$ North Pacific typhoons and found a median of $24$~hPa~day$^{-1}$.  The $75$th percentile, $42$~hPa~day$^{-1}$, was defined as the rapid intensification threshold. For North Atlantic  cyclones, the threshold was defined in terms of daily velocity increment \citep{kaplan03}.  Defining intensification rate via central pressure tendency has the merit of emphasizing that a low surface pressure is necessary to intensify the storm{\textquoteright}s circulation \citep{rodgers81}.

Since the atmosphere is approximately hydrostatic, in order for the surface pressure to drop, the mass of the atmospheric column must decrease. From this perspective, rapid intensification represents a remarkably subtle imbalance between the inflow $F_{\rm in}$ and outflow $F_{\rm out}$ of air in the storm core. 
Consider an axisymmetric cyclone with radial velocity $u_{\rm in}$ in the vicinity of the radius of maximum wind $r_m$ within the boundary layer of depth $h_{\rm in}$. How fast does the inflow 
$F_{\rm in} = 2 \rho_s u_{\rm in} h_{\rm in} /r_m$ (see Appendix A) recycle matter within the cylinder of radius $r_m$? The turnover time is given by
\begin{equation}\label{tau}
\tau_F = \frac{\rho_{s} H}{F_{\rm in}} = \frac{r_m H}{2 u_{\rm in} h_{\rm in}}, 
\end{equation}
where  $\rho_{s}$ is air density at the surface, $H$ is the exponential height of air density, and $\rho_s H$ (kg~m$^{-2}$) is the total air mass of the atmospheric column. For characteristic values $u_{\rm in} = 20$~m s$^{-1}$, $r_m=40$~km, $h_{\rm in}=1$~km, $H \sim 10$~km, and $\rho_{s}=1$~kg~m$^{-3}$, 
we obtain $\tau_F  = 10^4$~s~$\simeq 2.8$~hours.

Meanwhile with a typical maximum intensification rate of the order of $20$~hPa~day$^{-1}$, the entire column with surface pressure $~10^3$~hPa would have been depleted in $\tau_P = 50$~day~$\simeq 4 \times 10^6$~s. This means that, even during maximum intensification, the inflow and outflow of air into and from the hurricane core must coincide with each other with a high precision of a fraction of percent, $\tau_F/\tau_P = 0.002$. The outflow and inflow are apparently tightly {\textquotedblleft}slaved{\textquotedblright} to one another \citep{smith16}.

What determines this subtle imbalance between the outflow and inflow of atmospheric air that dictates how fast the storm intensifies? Is there a process  inherent to tropical cyclones that would be characterized by a relevant magnitude? Yes, there is such a process.  Tropical cyclones are well known for their extreme precipitation rates. In the North Atlantic hurricanes precipitation within $100$~km from the center reaches $P = 200$~kg~m$^{-2}$~day$^{-1}$ exceeding the local long-term climatological mean by a factor of forty\footnote{\normalsize
 Divided by density $\rho_l = 10^3$~kg~m$^{-3}$ of liquid water, precipitation $P$ (kg~m$^{-2}$~day$^{-1}$) is commonly measured in millimeters per unit time, e.g., 1~kg~m$^{-2}$~day$^{-1}$ corresponds to 1 mm~day$^{-1}$.} \citep{lonfat04,ar17,tu21}. If matter is removed from a hydrostatic air column at this rate, surface pressure will drop by $gP \simeq 20$~hPa~day$^{-1}$ ($g$ is the acceleration of gravity). This is close to the characteristic maximum intensification rate as established by \citet[][]{holliday79}. In other words, during maximum intensification the mean surface pressure tendency in the storm{\textquoteright}s core of radius $r$  is of the order of local precipitation times $g$ and can be written as (see Appendix A):
\beq\label{pt}
\frac{1}{g}\frac{\pt p}{\pt t} =F_{\rm in} - F_{\rm out} - P \sim - P.
\eeq

Despite it being the leading term in the mass budget for the storm core, the condensation mass sink appears to have been completely excluded from theoretical investigations of storm intensification. Even when {\it diagnosing} the central pressure tendency from the continuity equation,  the mass sink has recently been set to zero \citep[see Eq.~(1) and subsequent derivations of][]{sparks22a}. 
This situation appears to have been grounded in a long-standing confusion, which we here intend to clarify.

\citet{la04} estimated that precipitation in Hurricane Lili (2002) was sufficient to account for the storm{\textquoteright}s intensification rate and used that as a motivation to investigate the potentially significant impact of the precipitation mass sink on storm intensity. 
Since water vapor is a minor atmospheric constituent, in the continuity (mass conservation) equation the mass sink is a minor term that is often neglected. \citet{la04} restored the mass sink in the system of equations solved by a numerical model for Hurricane Lili (2002) and compared the output of thus modified model with the control (with no mass sink in the continuity equation). 

The authors concluded that the mass sink was not negligible but not dominant either (increasing the maximum drop of surface air pressure by a few hectopascals and maximum intensity by a few meters per second, Table~\ref{tab1}). However, this conclusion cannot be considered satisfactory for a reason that, surprisingly, has never been discussed: the mass  sink was present in the control simulation as well. \citeauthor{la04}{\textquoteright}s~\citeyearpar{la04}  Fig.~10c shows the relatively modest difference in the accumulated precipitation between the simulation with the mass sink included (MSNK) and the control (CTRL). But if CTRL had not had a mass sink, how could it have generated precipitation? Apparently, \citeauthor{la04}{\textquoteright}s~\citeyearpar{la04} control run, as well as control runs in other related studies \citep[e.g.,][]{qiu93,dool93}, did all include a mass sink.

\begin{table}[!ht]       
    \begin{minipage}[p]{1\textwidth}
    \caption{\normalsize Selected studies of the condensation mass sink in tropical storms, and their findings.}\label{tab1}
    \begin{threeparttable}
    \centering    
    \begin{tabular}{llcccl}
    \hline \hline 
      \multicolumn{1}{l}{Study}& Cyclone&\multicolumn{1}{c}{Mass} & \multicolumn{1}{c}{Condensation}   & \multicolumn{1}{c}{Maximum}\\ 
    \multicolumn{1}{c}{} &&\multicolumn{1}{c}{conservation$^*$}& \multicolumn{1}{c}{mass sink}   & \multicolumn{1}{c}{velocity (m~s$^{-1}$)}\\ [2ex]
    \hline
\citet{la04}  &Modeled Hurricane Lili (2002), run CTRL& No &  Non-zero & $49$\\    
                  &Modeled Hurricane Lili (2002), run MSNK                     & Yes & Non-zero & $57$\\    \hline
\citet[][]{bryan09a},  &Pseudoadiabatic, {\textquotedblleft}traditional equation set{\textquotedblright} & No &  Non-zero&      $104$\\    
 their Fig.~10 and Table~4                 &                Pseudoadiabatic, {\textquotedblleft}conservative equation set{\textquotedblright}    & Yes  & Non-zero & $104$\\    
                  &           Reversible, {\textquotedblleft}conservative equation set{\textquotedblright}         & Yes &   Zero&  $40$ \\    \hline
\citet[][]{wang20},  &Control    & Yes &  Non-zero& $82$\\    
 their Table~3                 &       Reversible             & Yes   & Zero&  $41$ \\    
\hline   \hline
    \end{tabular}
\begin{tablenotes}[para,flushleft]
$^*$ {\textquotedblleft}No{\textquotedblright} -- condensation mass sink is accounted for in the energy equation only; {\textquotedblleft}Yes{\textquotedblright} -- condensation mass sink is accounted for in both energy and continuity equations.
\end{tablenotes}
\end{threeparttable}
\end{minipage}
\end{table}

In a subsequent influential study, \citet[][p.~1775]{bryan09a} also stated that they studied the {\textquotedblleft}precipitation mass sink{\textquotedblright} and found  that it {\textquotedblleft}had a small effect on $\upsilon_{\rm max}${\textquotedblright}. This conclusion drawn from their Fig.~10 rests on the comparison of their {\textquotedblleft}traditional equation set{\textquotedblright} (where mass sink is excluded from the continuity equation) with the {\textquotedblleft}conservative equation set{\textquotedblright} (where mass sink is retained in the continuity equation)\footnote{\normalsize Prof. Kerry Emanuel, who reviewed the work of \citet{bryan09a}, in a letter published on a blog with the permission from all parties, explicitly pointed out that the simulations shown in Fig.~10 of \citet{bryan09a} were intended to examine the condensation mass sink, with the conclusion that it was {\textquotedblleft}not quantitatively large{\textquotedblright}, see \url{https://noconsensus.wordpress.com/2010/10/26/weight-of-water-and-wind-hurricane-pros-weigh-in/}, assessed 27 April 2023.}. However, both equation sets do simulate a mass sink, i.e., they do generate precipitation that is governed by the condensate terminal velocity $V_t>0$ as shown in  \citeauthor{bryan09a}{\textquoteright}s~\citeyearpar{bryan09a} Fig.~10. Their comparison, therefore, does not say anything about the actual role of the mass sink in storm dynamics (Table~\ref{tab1}).

Indeed, the mass sink term has a double presence in the full system of the equations of hydrodynamics for a moist atmosphere: in the energy conservation equation, where it governs the rate of latent heat release, and in the mass conservation (continuity) equation, where it contributes to the exact budget of the mass of moist air.  A common feature in many models has been to omit the mass sink in the latter, while retaining it in the former. However, due to the equation of state, changes of pressure, temperature and mass of water vapor are unambiguously related and emerge as a solution of the full system. Therefore, previous studies comparing model outputs with and without the mass sink in the continuity equation did not study the dynamic impact of the mass sink {\it per se}. Instead, they evaluated the impact of the violation of mass conservation in the model equations (see Table~\ref{tab1}, third column). {\it How the non-conservation of mass impacts the model solutions} (it does modestly) is by itself a legitimate question. But it is not equivalent to {\it how the mass sink impacts storm dynamics}.

To study the latter, one should include/exclude the mass sink into/from  {\it both} the continuity and energy conservation equations while preserving latent heat release in the latter. A straightforward way of switching the mass sink off is by putting terminal velocity $V_t = 0$. Then water vapor condenses and latent heat is realeased, but the condensate does not fall out  and travels together with the air. 
Such cyclones were termed {\it reversible} as they follow reversible thermodynamics. Governed exclusively by heat, they develop slowly and reach lower intensities. \citet[][their Table~4]{bryan09a} showed that varying $V_t$ from zero to infinity (all condensate is immediately removed, the so-called pseudoadiabatic cyclones) changes maximum intensity from $40$~m~s$^{-1}$ to over $100$~m~s$^{-1}$ for cyclones initiated in a buyoantly neutral environment (Table~\ref{tab1}). 

With a reference to \citet{em88}, \citet[][]{bryan09a} hypothesized that the presence of condensate at lower $V_t$ causes an excessive negative buoyancy and hence lower intensity.  However, \citet{mpi4-jas} showed that the impact of condensate lifting on storm intensity is minor. \citet[][p.~1784]{bryan09a} themselves implicitly admitted that the buoyancy explanation may not be fully satisfactory and, having noted that the change in $V_t$ can be relevant for predicting changes in intensity, relegated this problem to future studies. 

Fourteen years later, the puzzle of the low intensity of reversible cyclones has not been resolved despite several studies explicitly addressed their energetics \citep[e.g.,][]{wang20,wang21}.  \citet{wang20} demonstrated significant similarities in the dynamics of dry and reversible storms and their sharp contrast with real storms. As we shall argue, the fundamental reason for both the similarities and difference is the condensation mass sink.

This paper presents two lines of evidence. First, we use the Tropical Cyclone Extended Best Track Dataset (EBTRK) \citep[][]{demuth06} and the Tropical Rainfall Measuring Mission (TRMM) \citep{huffman07} to investigate the dependence between intensification rate and precipitation in North Atlantic cyclones. We show that Eq.~\eqref{pt} holds over a relevant range of conditions, with maximum precipitation setting the limit for maximum intensification rate. Second, we use a numerical Cloud Model 1 (CM1) \citep[][]{br02} to show that when the condensation mass sink is switched off,  and the storm is solely driven by sensible and latent heat, intensification rates become an order of magnitude lower than in the presence of the mass sink. We then discuss how the condensation mass sink presence changes the storm dynamics. In the concluding section, we put the condensation mass sink into a broader atmospheric perspective.

\section{\large Data and Methods}

\begin{table}[!ht]      
    \caption{\normalsize EBTRK data for the Atlantic cyclones in $1998$--$2015$: number of data entries, mean intensification rates, precipitation, radius of maximum precipitation, and radius of maximum wind.}\label{tab2}
    \begin{threeparttable}
    \centering    
    \begin{tabular}{lrccccccccc}
    \hline \hline 
Dataset & $n$	&	$I$	&	$gP_m$	&	$g\overline{P}(r_P)$	&		$gP(r_m)$	&	$r_P$	&		$r_m$ 	\\ 
 & 	&	hPa~day$^{-1}$	&	hPa~day$^{-1}$	&	hPa~day$^{-1}$	&		hPa~day$^{-1}$	&	km		&	km 	\\ [2ex] \hline
All intensifying storms                & $3492$	         &	$10$	&	$14$	&	$13$	&			&	$141$	&			\\
Intensifying storms with known $r_m$ &$2764$	&	$11$	&	$16$	&	$14$	&		$12$	&	$90$	&		$68$	\\
Lifetime maximum intensification & $266$	         &	$23$	&	$17$	&	$15$	&			&	$97$	&			\\
Lifetime maximum intensification with known $r_m$&      $213$	&	$25$	&	$18$	&		$17$	&	$13$	&		$83$	&	$65$	\\
\hline   \hline
    \end{tabular}
\end{threeparttable}
\end{table}

To analyze the dependence between intensification rate and precipitation we followed the approach of \citet{ar17}. We used the EBTRK dataset released on 27 July 2016 \citep{demuth06} and the 3-hourly TRMM 3B42 (version 7) (spatial resolution $0.25^{\rm o}$ latitude $\times$ $0.25^{\rm o}$ longitude).
EBTRK data are recorded every six hours. For the years $1998$--$2015$, for each $k$-th record in the EBTRK dataset (with $(k-1)$th and $(k+1)$th records referring to the same storm), we defined the intensification rate $I_k \equiv -2 (p_{k+1}-p_{k-1})$ (hPa~day$^{-1}$), where $p$ (hPa) is the minimum central pressure. This procedure yielded a total of $7848$ values of $I$, of which $1304$ were zero, $3052$ were negative (deintensifying storms) and  $3492$ were positive (intensifying storms). We defined lifetime maximum intensification rate $I_m$ as the maximum $I$ value for a given storm. If there were several records with maximum $I$ for a given storm, we took the earliest one. This yielded $269$ values of $I_m$ for $269$ storms, of which three were zero and not used in the analyses.

Using the TRMM data, for each $k$-th position of the storm center in EBTRK, we established the dependence of precipitation $P(r)$ on distance $r$ from the storm center, with $P(r_i)$ defined as the mean precipitation  in all grid cells with $25 (i-1) \le r \le 25 i$ (km),   $1 \le i \le  120$, $r_i \equiv 25\times (i-1)+12.5$~km. Examples of $P(r)$ distributions for individual storms are given in Fig.~11 of \citet{ar17}. Maximum value of thus obtained $P(r_i)$ and radius $r_i$ corresponding to this maximum were defined as $P_m \equiv P(r_P)$ and $r_P$, respectively, for the $k$-th record. Mean precipitation within the circle $r\le r_P$ was defined as $\overline{P}(r_P)$. Additionally, $P(r_i)$ for which $25 (i-1) \le r_m \le 25 i$ (km) was defined as precipitation $P(r_m)$  at the radius of maximum wind $r_m$ (Table~\ref{tab2}).

To enable numerical comparison between precipitation and intensification rates, we expressed precipitation in hPa~day$^{-1}$ by multiplying precipitation by $g$
(one mm of water per hour (multiplied by $\rho_l g$) is equivalent to $2.4$ hPa per day). Alternatively, one could express pressure tendency in mm~$\rm H_2O$~day$^{-1}$.

\begin{figure*}[h!]
\begin{minipage}[p]{0.6\textwidth}
\centerline{\includegraphics[width=1\textwidth,angle=0,clip]{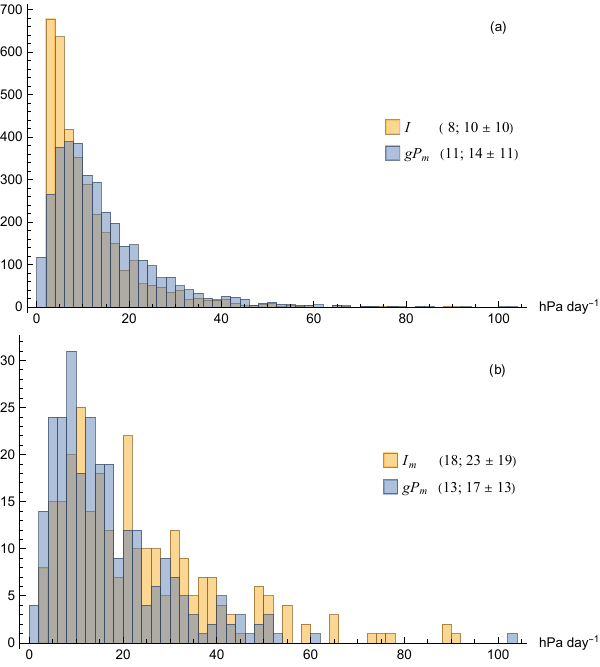}}
\end{minipage}
\caption{\normalsize
Frequency distribution of intensification rates $I$ (yellow) and concurrent maximum precipitation $gP_m$ (blue)
in North Atlantic cyclones in $1998$--$2015$. Figures in brackets are (median; mean $\pm$ standard deviation) (hPa~day$^{-1}$). (a) All $3492$ records for intensifying storms ($I > 0$); (b) Lifetime maximum intensification rates $I_m$ and concurrent maximum precipitation $gP_m$ for $266$ individual storms.
Two extreme values $I = I_m= 166$~hPa~day$^{-1}$ and $I= 128$~hPa~day$^{-1}$ recorded for hurricane Wilma $2005$ are not shown but included in the statistics.
}
\label{hist1}
\end{figure*}

For numerical modeling we used CM1 release 20.3 \citep{br02}. We used the default configuration for the axisymmetric runs (provided in the {\textquotedblleft}hurricane\_axisymmetric{\textquotedblright} directory
of the model configuration) with the following minimal modifications. We used $\rm ptype =6$ for the {\textquotedblleft}simple Rotunno-Emanuel (1987) water-only scheme{\textquotedblright} \citep{ro87}. We set terminal velocity $V_t = 0$ for reversible (REV) and $V_t = -10$ for pseudoadiabatic (PSE) runs. In the CM1 model, negative $V_t$ with $\rm ptype =6$ correspond to the pseudoadiabatic regime when the condensate is immediately removed from the atmosphere.

Additionally, we used a model run termed NOEV, where condensate did not fall out and the evaporation of cloud water in the atmosphere was switched off by changing the line {\textquotedblleft}IF(q3d(i,j,k,nqc).gt.1.0e-12 .or. q3d(i,j,k,nqv).gt.qvs)THEN{\textquotedblright} to
{\textquotedblleft}IF(q3d(i,j,k,nqv).gt.qvs)THEN{\textquotedblright} in the $\rm satadj$ subroutine.

As the initial temperature and moisture profiles, for REV and NOEV we used the reversible profile, and for PSE we used the control profile from  \citet[][their Fig.~1]{wang20} that both correspond to a surface air temperature $T_a = 296.15$~K. Sea surface temperature was set to  $T_s = 306.15$~K and $T_s = 301.15$~K for REV and NOEV and for PSE, respectively, such that the difference $\Delta T_s \equiv T_s - T_a$ was $10$~K for REV and NOEV and $5$~K for PSE.

We analyze the results of simulations for $400$ hours, which exceeds the lifetime of more than $95\%$ of real tropical cyclones \citep[][their Supplementary Fig.~2]{lee16}. All information necessary to replicate our simulations, as well as the $P(r)$ distributions for individual storm records, can be found at \url{https://doi.org/10.5281/zenodo.10577109}. 

\begin{figure*}[h!]
\begin{minipage}[p]{0.7\textwidth}
\centerline{\includegraphics{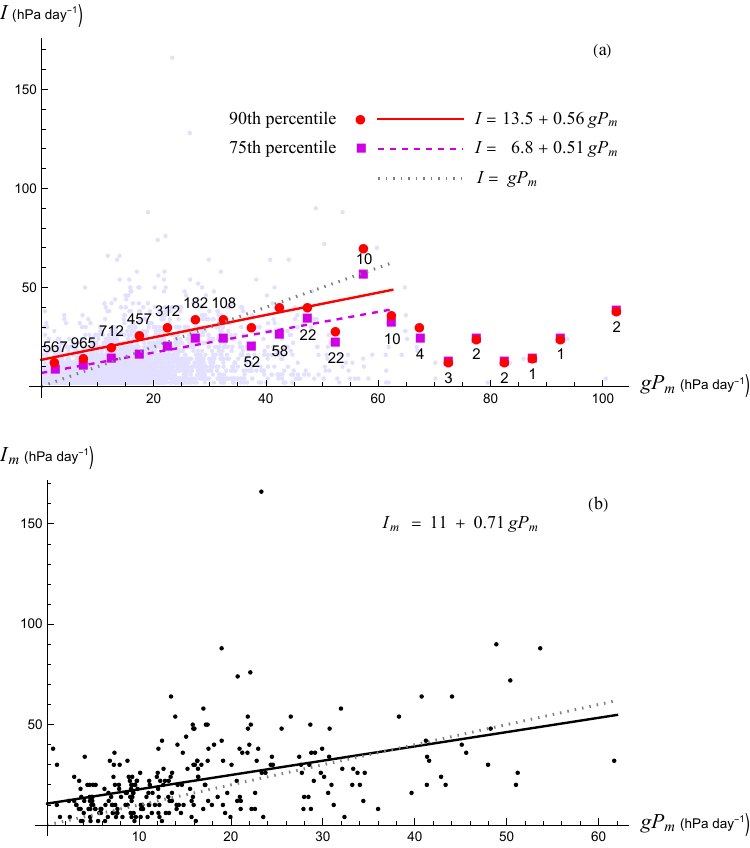}}
\end{minipage}
\caption{\normalsize
Dependence between intensification rates and maximum precipitation. (a) Linear regressions on $gP_m$ of the $90$th and $75$th percentiles of $I$ in $5$~hPa~day$^{-1}$ bins of $gP_m$ with $n\geq10$. Numbers $n$ of $I$ values in each bin are shown. Small dots are all $I$ values. (b) Linear regression of lifetime maximum intensification $I_m$ on $gP_m$. Dotted lines $I = gP_m$ (a) and $I_m = gP_m$ in (b) are shown for reference. In (b), one outlier (Hurricane Omar 2008 with  $gP_m =102$~hPa~day$^{-1}$ and $I_m = 38$~hPa~day$^{-1}$) is neither shown nor taken into account in the regression of $I_m$ on $gP_m$. 
}
\label{figIPm}
\end{figure*}

\section{\large Intensification and precipitation in North Atlantic tropical cyclones}

Intensification rates $I$ and maximum local precipitation $gP_m$ are on average close in magnitude being both of the order of $10$~hPa~day$^{-1}$ (Fig.~\ref{hist1}). Their frequency distributions 
differ in that  $I$ has a higher peak at lower values (Fig.~\ref{hist1}a): there are more cases of slowly intensifying storms than of intensifying storms with low precipitation. For lifetime maximum intensification rates $I_m$
this feature is not present, the means of $I_m$ and $gP_m$ are both close to $20$~hPa~day$^{-1}$ (Fig.~\ref{hist1}b).
The median $I_m = 18$~hPa~day$^{-1}$ for $266$ intensifying North Atlantic storms that happened in $1998$--$2015$ is comparable to the median $I_m = 24$~hPa~day$^{-1}$ for $305$ North Pacific storms that happened in $1956$--$1976$ \citep{holliday79}. Note, however, that our $I$ and $I_m$ were defined using a period of twelve hours rather than twenty four as in \citep{holliday79}.

The quantitative similarity between $I$ and $gP_m$ is not limited to their means. In $5$~hPa day$^{-1}$ bins of $gP_m$, the $90$th and $75$th percentiles of $I$ grow with $gP_m$ (Fig.~\ref{figIPm}a). Lifetime maximum intensification rates $I_m$ also increase with $gP_m$ (Fig.~\ref{figIPm}b). The intercepts at zero precipitation are around $10$~hPa~day$^{-1}$.

\begin{figure*}[h!]
\begin{minipage}[p]{0.95\textwidth}
\centerline{\includegraphics[width=1\textwidth,angle=0,clip]{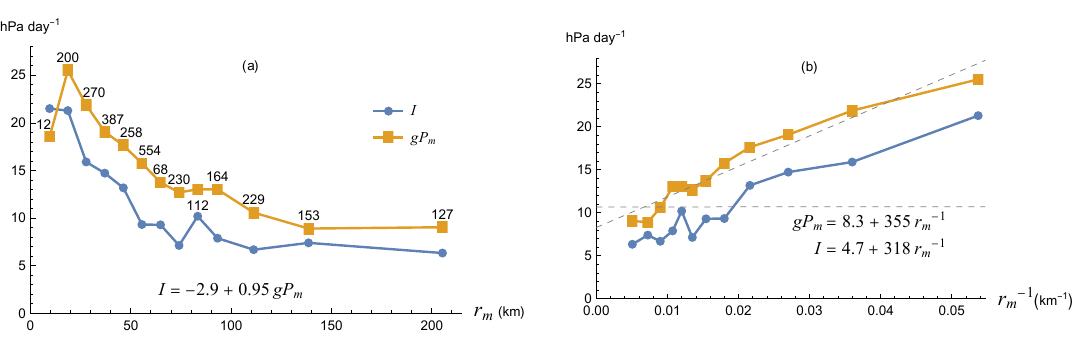}}
\end{minipage}
\caption{\normalsize
Dependence of intensification rate $I$ and maximum concurrent precipitation $gP_m$ on the radius of maximum wind (a) and inverse radius of maximum wind (b). Mean values of $I$ and $gP_m$ in each bin of $r_m$ are shown versus mean values of $r_m$ (a) and $r_m^{-1}$ (b) in each bin. Numbers $n$ of $I$ and $gP_m$ pairs of values in each bin are shown. Linear regression of the mean values of $I$ and $gP_m$ is shown in (a). In (b), the bin with the smallest $r_m$ and the smallest $n=12$ is neither shown nor taken into account in the regressions of $I$ and $gP_m$ on $r_m^{-1}$.
}
\label{figrad}
\end{figure*}

Figure~\ref{figrad} shows that both intensification rate and maximum concurrent precipitation depend similarly on the radius of maximum wind,
with $gP_m$ being at all but the smallest $r_m$ larger than $I$ by approximately $4$~hPa~day$^{-1}$.

We emphasize that we are not just dealing with a positive correlation between $I$ and $gP$, but with their numerical equivalence. Central pressure is diminishing approximately as it would if the inflow and outflow at the radius of maximum precipitation were exactly compensated, $F_{\rm in} = F_{\rm out}$, and surface pressure within this radius declined at a rate dictated by precipitation alone.  This simplified picture  certainly has its limitations, as surface pressure does not decline uniformly everywhere within the radius of maximum precipitation.

\begin{figure*}[h!]
\begin{minipage}[p]{0.9\textwidth}
\centerline{\includegraphics[width=1\textwidth,angle=0,clip]{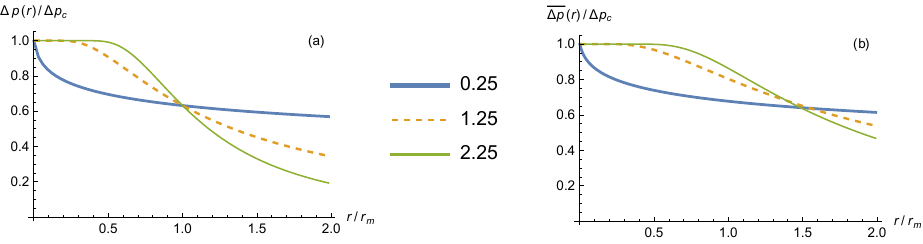}}
\end{minipage}
\caption{\normalsize
Dimensionless Holland{\textquoteright}s pressure profile $\Delta p(r)$, Eq.~\eqref{pH}, and the mean pressure drop $\overline{\Delta p}(r) = 2r^{-2} \int^r_0 \Delta p(r') r' dr'$  within a circle of a given radius for three values of $b = 0.25$, $1.25$, and $2.25$.
}
\label{figholl}
\end{figure*}

Using Holland{\textquoteright}s pressure profile
\beq\label{pH}
\Delta p(r) \equiv p(\infty) - p(r) = \Delta p_c \left[1-\exp\left\{-\left(\frac{r_m}{r}\right)^b \right\} \right] , 
\eeq
where $\Delta p(r) >0$ and $\Delta p_c > 0$ are the surface pressure drop at radius $r$ and the storm center $r = 0$, respectively,
$b \sim 1$ is a scaling parameter \citep[][their Eq.~(2)]{holland80,vickery08,holland10}. At the radius of maximum wind $r = r_m$, the pressure drop does not depend on $b$: $\Delta p_{m} \equiv \Delta p(r_m) = (1 -e^{-1}) \Delta p_c \simeq 0.6 \Delta p_c$ (Fig.~\ref{figholl}a).   
Likewise, the mean pressure drop $\overline{\Delta p}$ does not depend on $b$ at $r \simeq 1.5 r_m$  where it also constitutes about $0.6 \Delta p_c$ (Fig.~\ref{figholl}b).

If this relationship holds during intensification, one can expect that within the circle $r \le 1.5 r_m$  the mean pressure tendency  $I(1.5 r_m) \equiv  \pt \overline{\Delta p}(1.5 r_m)/\pt t$ will be by a factor of about $0.6$ smaller than $I \equiv \pt \Delta p_c/\pt t$, $I(1.5 r_m) \simeq 0.6 I$. 

The radius of maximum precipitation $r_P$ is close to $1.5 r_m$ in tropical cyclones of category $1$-$3$ and about $r_P \simeq 1.2 r_m$ in the more intense and in the less intense cyclones \citep[][their Fig.~2]{wang24}. If $I(1.5 r_m)$ is approximately determined by $g\overline{P}(r_P)$, then we can expect $g\overline{P}(r_P) \simeq 0.6 I_m$. Figure~\ref{IlPrmw} shows that the medians and means of $0.6 I_m$ and $g\overline{P}(r_P)$ are very close.
These spatial patterns require further study.

\begin{figure*}[h!]
\begin{minipage}[p]{0.7\textwidth}
\centerline{\includegraphics{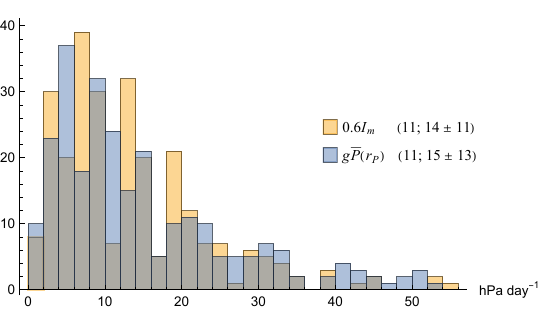}}
\end{minipage}
\caption{\normalsize
Frequency distribution of lifetime maximum intensification rates $I_m$ multiplied by a factor $0.6$ (yellow) and mean precipitation 
$g\overline{P}(r_P)$ within the radius $r_P$ of maximum precipitation (blue) in $266$ North Atlantic cyclones in $1998$--$2015$. Figures in brackets are (median; mean $\pm$ standard deviation) (hPa~day$^{-1}$). Two extreme values, $I_m= 166$~hPa~day$^{-1}$ for hurricane Wilma $2005$  and $g\overline{P}(r_P) =102$~hPa~day$^{-1}$ for Hurricane Omar $2008$, are not shown but taken into  account in the statistics.
}
\label{IlPrmw}
\end{figure*}

\section{\large Intensification of model storms without a mass sink}

While in real life latent heat release is not separable from the condensation mass sink, numerical models represent  a convenient tool to study the two effects separately. Our approach was not to do any tampering with the model parameters, but to use the default values of the CM1 model that, under realistic atmospheric conditions, would produce a realistic axisymmetric storm. Figure~\ref{figtime} shows that when the mass sink is switched off and the condensate does not fall out, the resulting reversible storm develops significantly more slowly than the storm with a condensation mass sink despite the greater temperature difference with the ocean in the former. In $400$ hours, the reversible storm develops a central pressure drop of $\Delta p_c = 10$~hPa and maximum velocity $\upsilon_{m} = 36$~m~s$^{-1}$ (and continues to intensify). The pseudoadiabatic storm develops $\Delta p_c = 128$~hPa and maximum velocity $\upsilon_{m} = 107$~m~s$^{-1}$ in $54$ hours and begins to deintensify slowly after $t = 100$~h \citep[a possible cause for this deintensification is the depletion of atmospheric moisture in the subsidence region, see][]{rizzi21}. By varying model setups, it is possible to make reversible storms develop more rigorously than the one shown in Fig.~\ref{figtime}, but a drastic difference in the intensification rates persists \citep[cf.~Fig.~2 of][]{wang20}. Moreover, storms without condensation mass sink increase their radius of maximum wind as they intensify (Fig.~\ref{figtime}c).

\begin{figure*}[h!]
\begin{minipage}[p]{0.5\textwidth}
\centerline{\includegraphics[width=0.8\textwidth,angle=0,clip]{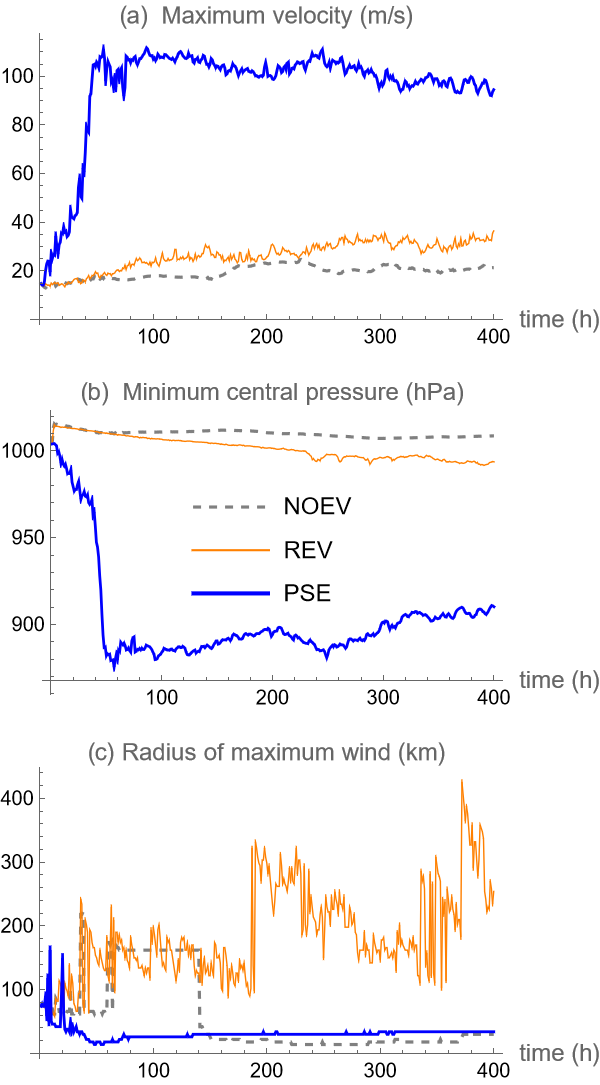}}
\end{minipage}
\caption{\normalsize
Intensification of a pseudoadabatic storm (PSE), reversible storm (REV) and a storm with evaporation in the air switched off (NOEV).
The initial temperature difference with the ocean is $10$~K in REV and NOEV and $5$~K in PSE.
}
\label{figtime}
\end{figure*}

Let us consider a simplified qualitative picture of this intensification (Fig.~\ref{figsch}a). The surface fluxes of water vapor and heat are parameterized as proportional  to the absolute wind speed following Eqs.~(34) and (35) of \citet{ro87}. Therefore, at the initial moment of time at the radius of maximum wind $r_m=80$~km of the initial tangential wind profile, the air is warming more intensely compared to the air at other radii. As this warmer air begins to rise,
there forms an outflow governed by the pressure surplus caused by the higher temperature in the upper troposphere\footnote{\normalsize Figure~\ref{figsch}a illustrates the irrelevance of the buoyancy argument as a possible explanation of the differences in the behavior of storms with and without a mass sink \citep[cf.][p.~1783]{bryan09a}. Buoyancy reflects the difference between local air density and a reference air density at the same pressure. In a reversible storm, the  condensate is present everywhere and its absolute amount does not matter. What matters is the extra $\Delta q_l$ of condensate mixing ratio in the region of ascent as compared to the region of subsidence. The relative decrease in buoyancy due to this extra condensate is equal to $\Delta q_l$, while the relative increase in buoyancy caused by condensation is $\Delta T/T = L \Delta q_l /(c_p T)$, where $L$ (J~kg$^{-1}$) is the latent heat of vaporization and $c_p$ (J~kg$^{-1}$~K$^{-1}$) is the specific heat capacity of air at constant pressure. Since at $T \simeq 300$~K we have $L/(c_p T) \sim 8$, the extra condensate makes only a minor negative impact on the buoyancy of the ascending air compared to the case when the condensate falls out.
}.

As the upper air flows away, surface pressure in the warmer region declines. This pressure drop causes an inflow in the lower atmosphere, which grows rapidly to match the outflow. The inflow is {\textquotedblleft}slaved{\textquotedblright} to the outflow as it is a function of the pressure drop created by the outflow. Accordingly, intensification continues as long as the outflow and the surface pressure drop it creates increase. Intensification is related to increasing temperature \citep{schubert82,emetal94}: the outflow grows as long as the ascending air warms ($\pt T/\pt t > 0$). As soon as thermal equilibrium is reached and $\pt T/\pt t = 0$, the intensification stops. Apparently, in the heat-driven storm shown in Fig.~\ref{figtime}, the hypothetical feedback between a decrease in the radius of maximum wind and an increase of latent heat release to cause $\pt T/\pt t > 0$ and drive further intensification (the so-called increased latent heat efficiency), is not realized \citep{schubert82,smith16b}.

\begin{figure*}[h!]
\begin{minipage}[p]{0.95\textwidth}
\centerline{\includegraphics[width=0.8\textwidth,angle=0,clip]{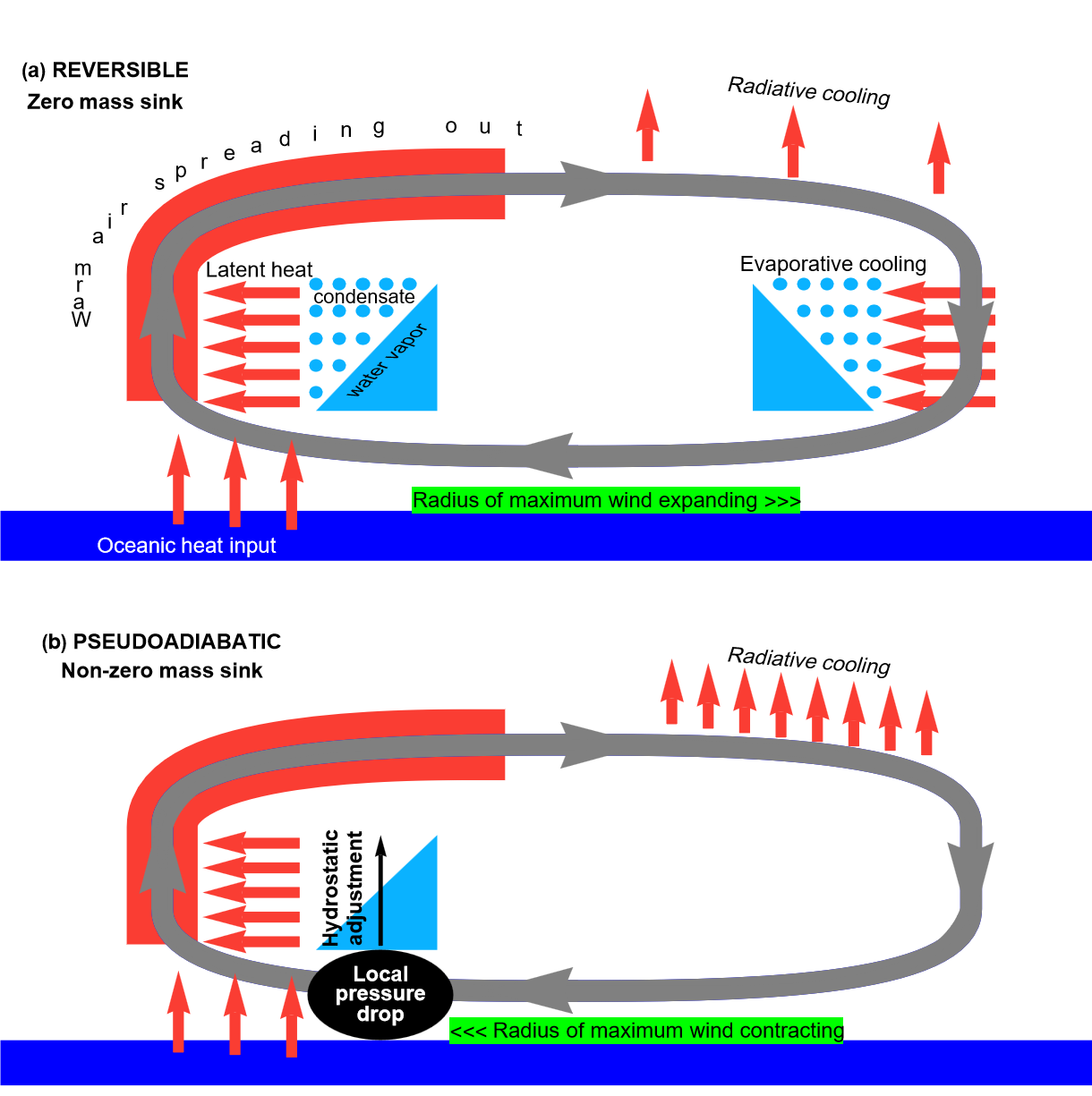}}
\end{minipage}
\caption{\normalsize
Key mechanisms at play during intensification of storms without (a) and with (b) condensation mass sink. In (b), since there is no condensate to evaporate, all latent heat released in the ascent region must be radiated to space in order for the air to descend.
}
\label{figsch}
\end{figure*}

What determines how slowly the storm intensifies? In this model setup, the atmosphere receives energy (in the form of latent and sensible heat) from the warmer ocean of constant temperature and loses energy via radiative cooling parameterized following Eq.~(30) of \citet{ro87}:
\beq\label{cool}
Q_r = \frac{\theta - \overline{\theta}}{\tau_r}.
\eeq
\noindent
This relaxes the potential temperature profile $\theta(z,r)$ to the initial state $\overline{\theta}$. The time scale $\tau_r$ is chosen from the condition that the resulting cooling should approximately balance the subsidence region in realistic cyclones (such that $Q_r \simeq 2$~K~day$^{-1}$).

Importantly, the descending air must lose all heat it has gained in the region of ascent. Otherwise it would end warmer at the surface than it started the ascent. The thermodynamic cycle would be characterized by negative work, as the air in the boundary layer would have to cool while moving from the subsidence region towards the region of ascent. This argument was comprehensively presented by \citet{goody03}. Accordingly, the subsidence must occur slowly enough for the air to dispatch all gained heat. But \citet{kieu09} showed that the intensification of tropical cyclones depends on how rapidly the secondary circulation intensifies.  Thus, because secondary circulation (vertical velocity in the subsidence region) is limited by radiative cooling, the storm cannot develop faster than allowed by the cooling parameterization.  

This limitation could be overcome, and the secondary circulation proceed more rapidly, if the storm core became more compact (the air can in principle ascend with an infinite velocity over an infinitely small area). But instead we can see that the radius of maximum wind grows (Fig.~\ref{figtime}c). It grows because the horizontal profile of the pressure drop at the surface is determined by the warm outflow which naturally tends to expand rather than contract \footnote{\normalsize \citet[][p.~2418]{lindzen87} expressed a related idea about latent heat being inefficient in driving local low-level motions as follows (our emphasis): {\textquotedblleft}\ldots flows generated by cumulus heating do not contribute effectively to low-level convergence (at least for time scales $>1$ week) because the cumulus heating peaks in the upper troposphere {\it and the forced motions decay away from the heating maximum{\textquotedblright}}.}. As a result, in sharp contrast with real cyclones that generally reduce their $r_m$ during intensification \citep{willoughby90,li21,wu21}, modeled cyclones without a mass sink increase their radius of maximum wind from the initial vortex with $r_m = 80$~km to $r_m \gtrsim 200$~km \citep[see Fig.~\ref{figtime}c and][their Fig.~2c]{wang20}.

\section{\large Intensification of cyclones with a mass sink}

\subsection{\large Latent heat handicap}

Let us now consider how condensation mass sink modifies the intensification process (Fig.~\ref{figsch}b). First, we note that in the subsidence region there is no condensate to evaporate. In the reversible storm, evaporative cooling in the descending air serves as a sink for the most part of the latent heat released in the region of ascent (Fig.~\ref{figsch}a). In the absence of this heat sink, all this latent heat must now be radiated to space for the air to descend. This poses an even stricter limitation on the vertical velocity of subsiding air, which, for the same upward mass flow, should be several times less than in the reversible storm.

As we have already noted, the only way a storm can rapidly intensify under the strict limitation imposed by the need to get rid of the latent heat in the slowly descending air is by becoming more compact in the region of ascent. \citet{wang21} expressed this idea by relating storm compactness to irreversibility (associated with the lack of evaporative cooling in the descending air). However, irreversibility by itself cannot make the storm more compact. An irreversible storm {\it must} be more compact if it is to develop. But it may not develop at all if there is no dynamic process that ensures its compaction.

To illustrate this simple idea, we introduced irreversibility by switching off the evaporation of condensate while keeping all condensate in the air (model run NOEV in Fig.~\ref{figtime}). In this case there is no storm development. The circulation is unable to dispose of a huge amount of latent heat which therefore serves as a break on its intensification. Condensation mass sink absent, the storm cannot get more compact whether it is irreversible or not.

\subsection{\large Outflow by centrifugal force}

Condensation mass sink apparently provides for the needed compaction. As the warmer air begins to rise and flow away from the cylinder of radius $r=r_m$, there is now a distinct additional process to remove mass from the column and lower surface pressure: precipitation. In contrast to the temperature-driven outflow that tends to equalize the temperature difference across the domain which results in storm expansion, the mass sink reduces surface pressure {\it locally} -- where precipitation actually occurs. 

As in all storms, the inflow is driven by the surface pressure drop. But the surface pressure drop is now driven not only by an increase in the outflow due to rising temperature, but also by the mass sink. For inflow and outflow to remain closely coupled, there must be an outflow process that correlates with the inflow driven by the mass sink. This outflow is caused by the centrifugal force, which ventilates away all the air that rises high enough. Indeed, in both models \citep{kurihara75,smith18,mpi4-jas} and observations \citep[][their~Fig.~7]{he18,he19}, the radial pressure gradient is positive in the upper outflow region (around $12$~km) and serves as a sink of kinetic energy generated in the boundary layer.

Whether the centrifugal force is able to ventilate the upper air can be assessed as follows. Kinetic energy per unit mass of air is $\upsilon_{m}^2/2$, where $\upsilon_{m}$ is the maximum tangential velocity. It is observed in the narrow inflow layer near the surface.  Applying Bernoulli{\textquoteright}s equation \citep[e.g.,][his~Eq.~(1)]{emanuel91} to the horizontal streamline, we can write
\beq
\label{B}
\frac{\upsilon_{m}^{2}}{2} = k \frac{\Delta p_m}{\rho_s}, 
\eeq
where $\rho_{s} = 1$~kg~m$^{-3}$ is surface air density, $\Delta p_m > 0$ is the surface pressure drop at the point of maximum wind and factor $k \lesssim 1$ reflects the kinetic energy loss due to friction. Taking into account that at the radius of maximum wind the surface air is in (approximate) gradient-wind balance:
\beq\label{gb}
\frac{\pt p}{\pt r} \Big|_{r = r_m} = \rho \frac{\upsilon_{m}^2}{r_m},
\eeq
using Holland{\textquoteright}s pressure profile, Eq.~\eqref{pH}, and solving Eqs.~\eqref{B} and \eqref{gb} for $r = r_m$ and $k=k(b)$, we obtain \citep[cf.][their Eq.~(4)]{vickery08}:
\beq\label{k}
k = \frac{b}{2 (e - 1)}.
\eeq
For the range of $0.25 \le b \le 2.25$ as per \citeauthor{holland80}{\textquoteright}s~\citeyearpar{holland80} Fig.~2 we obtain $0.07 \le k \le 0.65$. With a mean $b \simeq 1.25$ \citep{vickery08}, we have $k \simeq 0.4$. This means that more than one third of the potential energy of the pressure drop along the streamline from the storms outskirts to the radius of maximum wind is converted to kinetic energy\footnote{\normalsize For a {\textquotedblleft}typical{\textquotedblright} tropical cyclone that develops $\upsilon_{m} = 60$~m~s$^{-1}$ and $\Delta p_c = 75$~hPa (from $1005$ to $930$~hPa) \citep[][Fig.~2]{schubert82}, we have $k = 0.4$ from Eqs.~\eqref{pH} and \eqref{B}.  For tropical cyclone Hato with $\Delta p \simeq 17$~hPa and $\upsilon \simeq 35$~m~s$^{-1}$ in the vicinity of the storm center \citep[][their Figs.~3, 5, and 7]{he19}, we obtain $k = 0.4$ directly from Eq.~\eqref{B}. For Hurricane Isabel (2003) with $\upsilon_{m} = 78$~m~s$^{-1}$ \citep[their Fig.~4a]{montgomery06} and $\Delta p_m \simeq 50$~hPa \citep[][their~Fig.~4]{aberson06} we have $k=0.6$ from Eq.~\eqref{B} (note that in this storm the wind was supergradient).
For PSE simulation, maximum $k$ during intensification was $k=0.76$ at $t = 21$~h.}. It is a big proportion. For Hadley cells, for example, with a typical pressure difference  $\Delta p = 10$~hPa and maximum velocity $\upsilon = 7$~m~s$^{-1}$, we would have $k \sim 0.02$.

The value $|\Delta p(z)|/\rho(z)$, where $\rho(z)$ is local air density and $\Delta p(z) \equiv p(\infty,z) - p(r,z)$ is the pressure difference between the unperturbed external environment and the local point, is equal to the minimum kinetic energy required to move air against this pressure difference from 
the eyewall to the external environment. If this energy is less than $\upsilon_{m}^2/2$,  the outflow of air by the centrifugal force is energetically permitted. Maximum possible outflow $U(z)\rho(z)$ is obtained from the condition that all kinetic energy is spent on the radial motion with velocity $U(z)$:
\beq\label{urho}
U(z) = \sqrt{2 k \frac{\Delta p_m}{\rho_s} - 2\frac{\Delta p(z)}{\rho(z)} }.
\eeq

\begin{figure*}[h!]
\begin{minipage}[p]{1\textwidth}
\centerline{\includegraphics[width=1\textwidth,angle=0,clip]{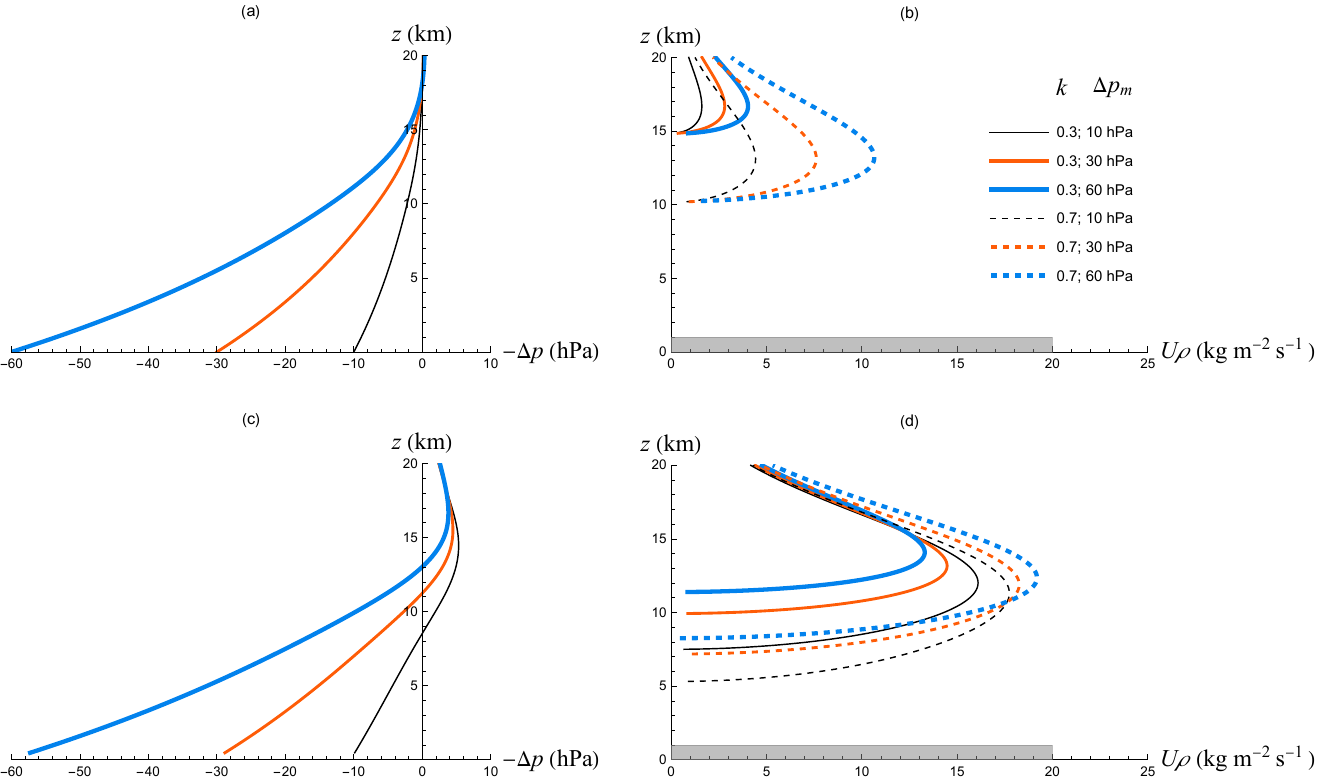}}
\end{minipage}
\caption{\normalsize
Vertical pressure differences $\Delta p(z) \equiv p_{0}(z)- p(z) $ (a,c) and the corresponding profiles $U(z)\rho(z)$  (b,d) calculated from Eq.~\eqref{urho}, 
where $p(z)$ is the vertical pressure of moist saturated air columns with a pseudoadiabatic lapse rate, surface air temperature $T_s = 298$~K and a surface pressure deficit of $\Delta p_m = 10$, $30$ and $60$~hPa compared to the reference surface pressure $p_{0}(0) = 10^3$~hPa. In (a), the reference pressure profile is a moist saturated air column with a pseudoadiabatic lapse rate and $T_s = 298$~K; in (b) it is an air column with $T_s = 300$~K and relative humidity $80\%$ at the surface; with dry adiabatic lapse rate up to the dew point at $z = 474$~m and a moist pseudoadiabatic profile above.
}
\label{figprof}
\end{figure*}

Figure~\ref{figprof}a,b describes a case when both the eyewall and the external environment have a moist pseudoadiabatic vertical temperature profile
with the same surface temperature $T_s = 298$~K, but the eyewall has a surface pressure deficit of $\Delta p_m = 10$, $30$ or $60$ hPa compared to
the external environment. In this case, the air pressure in the eyewall is lower than in the external environment at all heights below approximately $20$~km (Fig.~\ref{figprof}a). The gray bar in Fig.~\ref{figprof}b,d illustrates a typical magnitude of the low-level inflow $u_{\rm in} \rho_{s} h_{\rm in}$:  height of the inflow layer $h_{\rm in} =1$~km, horizontal velocity $u_{\rm in} = 20$~m~s$^{-1}$, and air density $\rho_s = 1$~kg~m$^{-3}$.

For an equal outflow to be energetically permitted, the area to the left of the $U(z)\rho(z)$ curves in Fig.~\ref{figprof}b,d should be equal to or greater than that of the gray bar. Figure \ref{figprof}a,b indicates that at the highest $k = 0.7$ the centrifugal force is able to ensure an adequate outflow between $10$ and $15$~km. This is where the outflow predominantly occurs in both real and modeled cyclones \citep{miller58,frank77,smith18}. For a given $k$, this energetically permitted outflow increases with increasing pressure deficit at the surface $\Delta p_m$. At lower $k = 0.3$ (less kinetic energy available for a given $\Delta p_m$) the air has to rise five kilometers higher and still there does not seem to be enough kinetic energy for it to be ventilated by the centrifugal force alone.

Figure~\ref{figprof}c,d shows a case where the eyewall has a small pressure surplus aloft not exceeding $2$~hPa between $10$ and $20$~km similar to the height-resolving pressure profiles for tropical cyclones established by \citet[][their Fig.~7]{he19}. The external environment is represented by an atmospheric
column with $T_s = 300$~K (i.e., it is by $2$~K warmer at the surface than the eyewall) but is unsaturated at the surface at $80\%$ relative humidity.
Even as the eyewall is colder at the surface (which may reflect the adiabatic expansion of inflowing air), its greater moisture content allows for a higher
temperature aloft that is responsible for the pressure surplus shown in Fig.~\ref{figprof}c. Accordingly, the energetically permitted outflow  increases, cf. Fig.~\ref{figprof}b,d. Now even at a lower $k=0.3$ the air can be ventilated if it rises to about $10$~km. Note that in this case for a given $k$ the higher the surface pressure deficit, the higher the air must rise to be ventilated. For example, with $k=0.7$ and $\Delta p_m=10$~hPa, the air can be ventilated from above $5$~km, while with $\Delta p_m=60$~hPa it must rise to above $8$~km. Figure~\ref{figdpevol} shows that during the storm development the pressure surplus aloft can be present at early stages but disappear later as the storm deepens.

\begin{figure*}[h!]
\begin{minipage}[p]{0.7\textwidth}
\centerline{\includegraphics[width=1\textwidth,angle=0,clip]{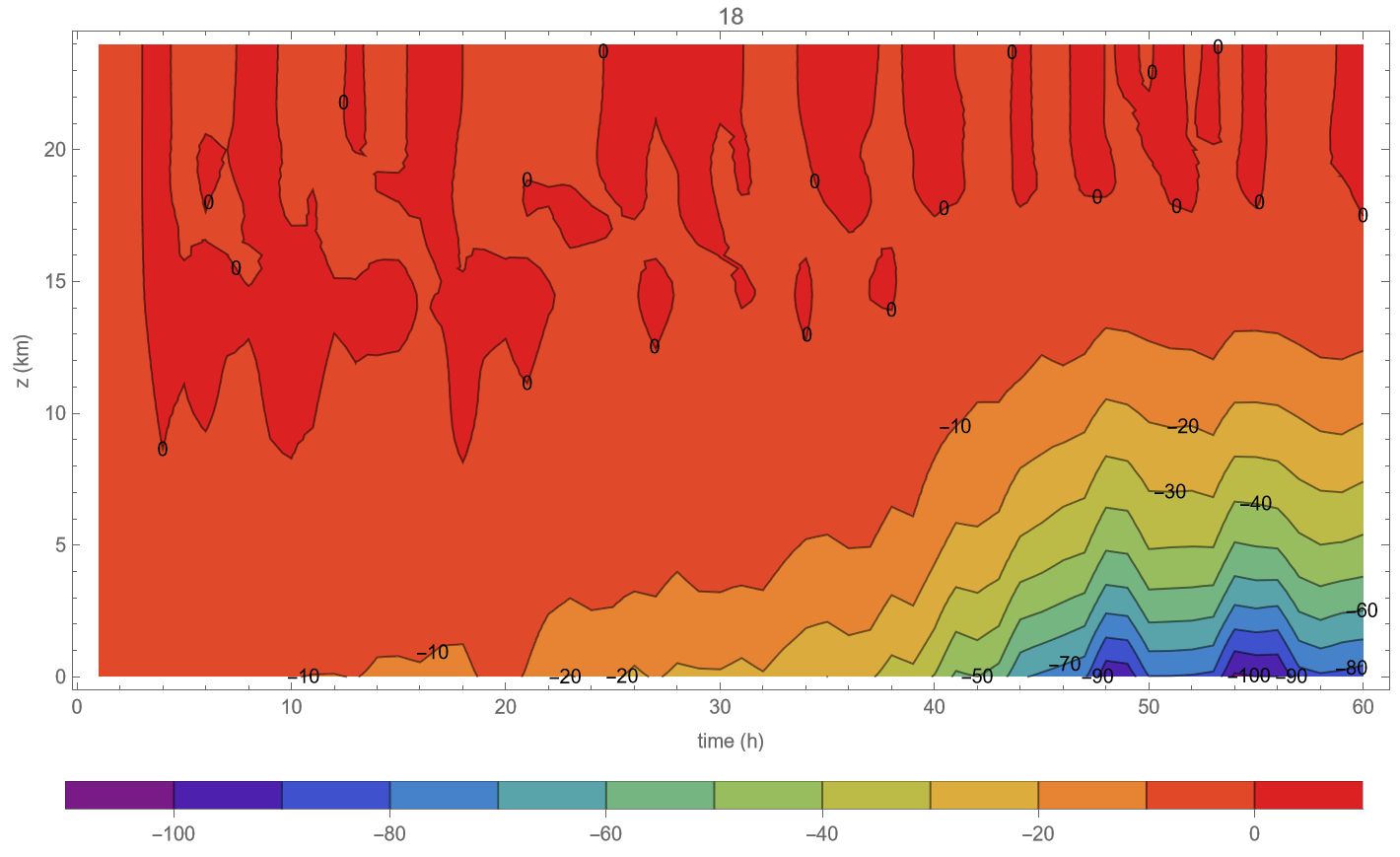}}
\end{minipage}
\caption{\normalsize
Time evolution of the difference $\Delta p(z)$ (hPa) between the vertical pressure profile at the radius of maximum wind and the outermost model radius in the PSE run during the first sixty hours of development. Note a small pressure surplus ($\lesssim 1$~hPa) in the upper troposphere at early stages of development and its disappearance later.
}
\label{figdpevol}
\end{figure*}

\subsection{\large Condensation, pressure adjustment and storm (de)-intensification}
\label{adj}

Figure~\ref{figprof} suggests that if the air rises sufficiently high, there is no problem for it to be ventilated from the eyewall under realistic atmospheric conditions.  However, if the ascending motion is compromised, so will be the outflow. From this position -- whatever rises flows away, and whatever does not rise does not flow away -- we can conceptualize the role of pressure adjustment that occurs during condensation in generating negative and positive pressure tendencies (Fig.~\ref{figadj}).

\begin{figure*}[h!]
\begin{minipage}[p]{0.7\textwidth}
\centerline{\includegraphics[width=1\textwidth,angle=0,clip]{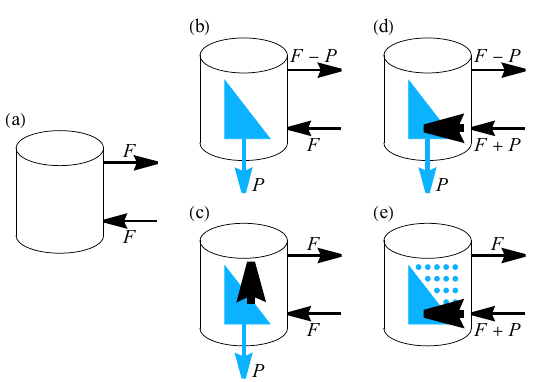}}
\end{minipage}
\caption{\normalsize
The role of pressure adjustment accompanying condensation in generating pressure tendencies. The triangle symbolizes the non-equilibrium vertical distribution of the partial pressure of water vapor that results from condensation; $P$ is precipitation, $F$ is the absolute steady-state magnitude of the inflow (the lower thin horizontal arrow) from the eyewall (cylinder) and outflow $F_{\rm out}$ (the upper thin horizontal arrow)  that are assumed to be balanced in the absence of the condensation (a).  Thick black arrows indicate the vertical (hydrostatic) pressure adjustment in (c) and the horizontal pressure adjustment in (d) and (e). The pressure tendency $(1/g)\pt p/\pt t = F_{\rm in} - F_{\rm out} - P$ is equal to zero in (a) and (b), to $-P$ in (c) and to $+P$ in (d) and (e), see text for details.
}
\label{figadj}
\end{figure*}

We consider condensation mass sink as a perturbation to the steady-state heat-driven circulation where the inflow and outflow are exactly balanced, $F_{\rm in} = F_{\rm out} = F$ and $\pt p/\pt t = 0$ (Fig.~\ref{figadj}a). A salient feature of the condensation mass sink is the formation of a strongly non-equilibrium
vertical gradient of the partial pressure of water vapor.  In a hypothetical case of no pressure adjustment to accompany condensation, this non-equilibrium would reduce the outflow by the magnitude of precipitating moisture, $F_{\rm out} = F - P$. Instead of rising to  be ventilated, water vapor would be exiting the column by condensation and precipitation. As a result, the pressure tendency equal to  $(1/g) \pt p/\pt t = F_{\rm in} - F_{\rm out} - P = 0$ would remain zero (Fig.~\ref{figadj}b). 

Hydrostatic adjustment redistributes the air upwards such that the void  caused by condensation is occupied by the air from below. If the hydrostatic adjustment pushes the air sufficiently high up \citep[this vertical scale is governed by the mean precipitation height $H_P$,][]{jas13}, then the outflow can regain its steady-state unperturbed value $F_{\rm out} = F$. Instead of rising to  be ventilated, water vapor condenses and precipitates, but its place is occupied by the dry air
pushed up by the hydrostatic adjustment. In this case, the pressure tendency is negative,  $(1/g) \pt p/\pt t = F_{\rm in} - F_{\rm out} - P = -P$ (Fig.~\ref{figadj}c). The storm intensifies.

But it is also possible, especially if the eyewall is narrow and/or the mean precipitation height is low, that the predominant direction of the pressure adjustment
will be in the horizontal plane (Fig.~\ref{figadj}d). This will increase the inflow $F_{\rm in}$ to $F+P$ \citep[the absolute magnitude of radial velocity will grow by the so-called barycentric velocity, see discussion by][]{acp13} and can cause a positive pressure tendency of the same magnitude,  $(1/g) \pt p/\pt t =  F_{\rm in} - F_{\rm out} - P = +P$ (Fig.~\ref{figadj}d). Horizontal adjustment can be compared to the {\textquotedblleft}back pressure{\textquotedblright} as formulated by \citet{lindzen87} but probably on a shorter timescale. When the condensate does not fall out, the horizontal pressure adjustment is possible, while the vertical adjustment is not (Fig.~\ref{figadj}e). This illustrates why it is difficult for the reversible storm to intensify.

This ability of water vapor condensation to generate both positive and negative pressure tendencies of (at maximum) the same absolute magnitude determined by precipitation, depending on the geometry of the pressure adjustment, is quite remarkable. It can provide a clue to the very peculiar pattern of intensification and de-intensification rates in real storms having approximately the same mean magnitudes (Fig.~\ref{figdeint}). They also depend similarly on the radius of maximum wind \citep{sparks22b} and by inference on precipitation (cf. Fig.~\ref{figrad}).

\begin{figure*}[h!]
\begin{minipage}[p]{0.6\textwidth}
\centerline{\includegraphics[width=1\textwidth,angle=0,clip]{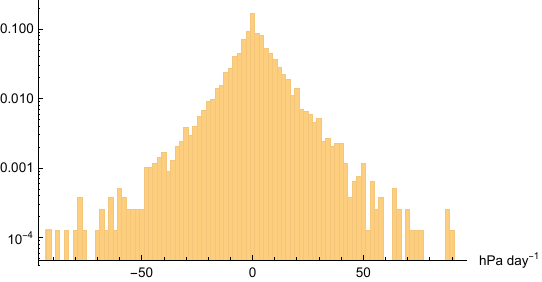}}
\end{minipage}
\caption{\normalsize
Frequency distribution of all $7848$ values of intensification rates including both intensifying (positive values) and de-intensifying (negative values) storms,
mean value is $\overline{I} = 0.7 \pm 13$~hPa~day$^{-1}$. Six outliers with $-112$~hPa~day$^{-1}$ $\le I \le -96$~hPa~day$^{-1}$ and two outliers with $I = 166$~hPa~day$^{-1}$ and $I=128$~hPa~day$^{-1}$ are not shown.
}
\label{figdeint}
\end{figure*}

This is peculiar, because model storms do not de-intensify at the same rate they intensify but tend to reach a steady state \citep[e.g., see discussion by][]{rizzi21}. This agrees with the idea that such a state is a stable attractor \citep[][]{kieu2015}. Indeed, if intensification of tropical storms is governed by self-induced dynamics characterized by its own timescale that is comparable to the storm{\textquoteright}s lifetime \citep[e.g., as shown in Fig.~2 of][]{schubert82},
why would their de-intensification occur at a similar rate?  The concept presented in Fig.~\ref{figadj} allows one to hypothesize that the intensification and de-intensification of the storm can be, in the zeroth approximation, viewed as a pulsation where the vertical pressure adjustment dominates during intensification and the horizontal adjustment during de-intensification. This said, we acknowledge the complexity of these matters and previous discussions \citep[e.g.,][]{kowch15,lee16,sparks22b} and emphasize the need for further studies.

How could the switch between the two regimes occur? The mean condensation height for complete condensation in the tropical atmosphere is about $6$~km \citep{jas13}. As Figure~\ref{figprof}d shows, with increasing surface pressure deficit in the eyewall $\Delta p_m$, height of the lower boundary of the outflow layer increases. As the storm deepens, when the lower boundary of the outflow significantly  exceeds the condensation height, we can hypothesize that hydrostatic adjustment will be unable to push the air high enough to keep the outflow unperturbed. Thus as the cyclone intensifies, on the one hand, the radius of maximum wind $r_m$ decreases and precipitation increases, and this enhances intensification. On the other hand, as the surface pressure deficit in the eyewall becomes too high,  the difference between precipitation height $H_P$ and the lower boundary of the outflow layer increases, and the intensification can stop. This may explain the fact that high-intensity cyclones like hurricanes of 3-5 category undergo rapid intensification several times less often than the less intense storms \citep{wang24}.

When the horizontal adjustment becomes dominant over the hydrostatic adjustment is a difficult theoretical question \citep{chagnon05,spengler11}. However, it is clear that the horizontal scale of the maximum precipitation area should play a role. As the storm becomes more compact and the eyewall width becomes less than $H_P$, horizontal adjustment should come into play and de-intensification commence. This can determine the minimal eyewall radius (of a few kilometers) that the most intense storms can attain.

On the other hand, the more complete condensation, the higher the mean precipitaiton height \citep[][their Fig.~1]{jas13}. This may explain why high convective bursts are known to be associated with rapid intensification \citep[e.g.,][]{chen13}. If $H_P$ grows abruptly and becomes higher than the lower boundary 
of the outflow layer, this can initiate rapid intensification. The subtle interplay between $H_P$, $\Delta p_m$, $r_m$ and $r_P$ that can cause the switch between intensification and de-intensification makes prediction of rapid intensification a challenging matter \citep[cf. "the subtle vertical structure issues" sensu][p.~1182]{masunaga2020}.

\section{\large Discussion and conclusions}

We have shown that the lifetime maximum intensification rates in North Atlantic tropical cyclones approximately coincide with local maximum precipitation and that turning off precipitation in the CM1 model greatly reduces the intensification rates. If one is not ready to immediately accept a major role for the condensation mass sink in storm dynamics, one faces the challenge of providing alternative explanations.  For example, if tropical cyclones intensify by WISHE (wind-induced surface heat exchange) or by vortical hot towers \citep[][]{montgomery09},  what clues do the respective concepts provide to anticipate that reversible cyclones (i.e., those with maximum thermodynamic efficiency) will not develop at all when the temperature difference between the ocean and the air is less than $5$~K \citep{wang20}, and when they will slowly develop, they will expand rather than contract? Understanding why storms do not intensify can be inseparable from understanding why they do.

The correspondence between intensification rate and precipitation is not the first indication of condensation mass sink possibly playing a major role in tropical storm dynamics. \citet{hess07} proposed that the non-equilibrium vertical distribution of water vapor partial pressure, which is compressed sixfold as compared to the hydrostatic distribution, represents a source of potential energy to drive winds.  This proposition (termed condensation-induced atmospheric dynamics) was later elaborated and found to be relevant in the context of hurricanes and tornadoes \citep{pla09b,pla11,tor11,jetp12,pla15}. In particular, the scale for the maximum velocity $\upsilon_{m}$ is given by $\rho \upsilon_{m}^2/2 = p_{\upsilon}$, where $\rho$ is air density and $p_{\upsilon}$ is partial pressure of water vapor that is interpreted as the maximum available potential energy. For typical temperatures $T=300$~K and relative humidity $80\%$ we have $p_{\upsilon} \simeq  35$~hPa which, with $\rho = 1$~kg~m$^{-3}$,  gives $\upsilon_{m} \simeq 80$~m~s$^{-1}$ corresponding to a category 5 hurricane.

Furthermore, if condensation of water vapor releases $p_{\upsilon}/N_{\upsilon}  = RT$~(J~mol$^{-1}$) of potential energy ($R=8.3$~J~mol$^{-1}$~K$^{-1}$ is the universal gas constant, $N_{\upsilon}= \rho_{\upsilon}/M_{\upsilon}$ is the molar density of water vapor), which is converted to kinetic energy and locally dissipates, then the local dissipation rate should be given by $PRT/M_{\upsilon}$ \citep{pla11,pla15}. For $P = 200$~kg~m$^{-2}$~day$^{-1}$ this gives $320$~W~m$^{-2}$. This coincides with the independently estimated dissipation rate $\rho C_D \upsilon_{m}^3 \simeq 300$~W~m$^{-2}$ of an average  Atlantic hurricane with  $\upsilon_{m} = 50$~m~s$^{-1}$, where $C_D = 2\times 10^{-3}$ is a characteristic surface drag coefficient  \citep{emanuel99}. Notably, for {\it global} precipitation $P = 1$~m~year$^{-1}$, the same expression $PRT/M_{\upsilon} = 4$~W~m$^{-2}$ yields a magnitude that is within $30\%$ of the {\it global} wind power, $3$~W~m$^{-2}$, independently estimated from observed wind speeds and pressure gradients \citep{jas13,arxiv17}.

We believe that the new evidence in favor of a major role of the condensation mass sink in storm dynamics that we have presented justifies  more attention to those arguments. We will conclude our discussion by placing the condensation mass sink into a broader atmospheric context.

\citet{kieu04,kieu06} showed that a realistic analytical solution for hurricane intensification could be obtained by postulating a positive feedback between the rate of latent heat release and vertical velocity. This is equivalent to assuming that  the faster latent heat is released in the rising air, the warmer the atmospheric column, the greater the buoyancy, and as a result the air rises even faster  \citep[see][his Eq.~1.5']{kieu04}.  However, \citet{emetal94}, see also discussion by \citet{stevens97} and \citet{emanuel97}, pointed out that just because moisture condenses faster, it does not follow that the air gets  warmer. In a subsequent publication, \citet{kieu09} showed that an analytical solution for hurricane intensification could be obtained by postulating an exponential growth of the vertical velocity, but at this point they  already did not emphasize that this assumption could actually reflect a (problematic) positive feedback between latent heat release  and vertical velocity [cf. Eq.~(1.7) of \citet{kieu04} and Eq.~(6) of \citet{kieu09}].

A certain lack of clarity on this point apparently persists in the literature. For example, \citet[][p.~147]{chen13} opined that {\textquotedblleft}clearly, the more intense the warm core,  the greater will be the hydrostatically induced surface pressure falls{\textquotedblright}, with the core intensity quantified as the updraft velocity. This is despite \citet[][p.~1140]{emetal94} warned that {\textquotedblleft}disturbance growth requires a positive correlation of heating and temperature, and perturbations of the latter  are usually very small, so that knowledge of the heating may be of little conceptual or predictive value.{\textquotedblright}

Some clues are provided by the earlier discussions of circulations driven by external differential heating or by latent heat release \citep[e.g.,][]{lindzen87,neelin89,an11}. External differential heating leads to a local increase in atmospheric scale height and vertical motion. There is a positive relationship: 
the greater the external heating, the greater the vertical motion due to increased temperature. As for latent heating, the positive relationship is different: the greater the vertical motion, the greater the latent heat release. This relationship has a different physical cause: it exists because the temperature gradient in the troposphere is approximately vertical, so condensation occurs predominantly as the air rises and cools.  In neither case there is a {\it positive  feedback}. In the first case, greater vertical velocity does not increase external differential heating (i.e., rising air does not make the sun brighter). In the second case, faster release of latent heat does not increase vertical velocity. But if the difference between latent heating and external heating is overlooked, it might appear that a positive feedback does exist.

\citet{Levermann2009,Levermann2016} assumed a positive feedback between latent heat release and air motion to explain the enigmatic abrupt changes in the monsoon circulations of the past climates, but \citet{Boos2016b}, see also \citet{Boos2016a}, clarified, with a reference to \citet{emetal94}, that this assumption was not plausible. More recently, see \citet[][their Eq.~(2)]{vallis20}, \citet{davison22} and \citet{vallis22}, showed that the same positive feedback between latent heat release and atmospheric warming can generate atmospheric patterns resembling the Madden-Julian Oscillation, a phenomenon that, like the rapid intensification of tropical storms, still defies a theoretical explanation.

Overall, it appears that positive feedback between condensation and atmospheric motion could be useful in elucidating several important problems in atmospheric science \citep[besides tropical storms, monsoons
and the Madden-Julian Oscillation, see also][]{masunaga2020}. However, theoretical studies in this direction are currently precluded by the valid objection that latent heat does not provide such feedback  \citep{emetal94,Boos2016b}. But the condensation mass sink does. With the pressure tendency proportional to precipitation, it should be possible to show that the derivations that previously assumed a positive feedback between latent heat release and vertical velocity will largely retain their form but acquire a different, and plausible, physical meaning. This is immediately clear in the case of the shallow-water equations like Eq.~(2) of \citet{vallis20} which, in the framework of condensation-induced atmospheric dynamics, would be equivalent to $(1/g)\pt p/\pt t = -P$ (radiative cooling increasing the pressure tendency can play the role of horizontal pressure adjustment, cf. Fig.~\ref{figadj}d).

The difference in intensification rates and maximum intensities of storms with and without the condensation mass sink shown in Fig.~\ref{figtime} can depend on model settings and especially on turbulence parameterization. In the CM1 model, turbulence parameters are determined from the condition that the resulting storm should be realistic. In the simulations of \citet{wang20}, who used a 3D model and different turbulence parameters than the default axisymmetric setup used in our study, the dry and reversible storms developed more rigorously than  the reversible storm shown in Fig.~\ref{figtime}. Using an older CM1 version and $\Delta T_s = 13$~K,  \citet{rizzi21} could obtain a dry storm that developed at a comparable rate with the control with a mass sink. Another observation is that storm intensity increases with increasing spatial resolution, both in CM1 \citep{bryan09a} and in global climate models \citep{daloz15}.  \citet{vallis20} likewise noted that for the self-sustained circulation mode to be preserved at shorter time steps, the horizontal resolution had to be increased (the grid size reduced). Conversely, \citet{lindzen87} noted that moisture convergence that was too high at a higher resolution came closer to observations when the grid size was increased.

Since efforts to increase the spatial resolution of global climate models go hand in hand with re-formulation of turbulence, the question is whether the right dynamics is preserved by these procedures. \citet{vallis22} pointed out  that condensation is a rapid process occurring on a shorter time and a smaller spatial scale than resolved by any model. Given that the time and spatial scales at which relevant dynamics arise remain unknown, not every combination of model parameters will be able to capture it. 

Conceivably, there can be model setups (unknown combinations of grid size, time steps, turbulence parameters) that, to a certain degree, suppress condensation-induced dynamics and enhance heat-driven dynamics, or vice versa. Such model setups will have different behaviors and generate different predictions for atmospheric phenomena involving condensation. For example, there are first indications that higher resolution models predict a smaller decline in the Amazon rainfall upon deforestation because of enhanced air convergence due to more sensible heat over deforested land \citep{yoon23}. This could be an example of  an artificially enhanced heat-driven circulation as compared to the model with a lower resolution. Indeed, recent studies suggest that global climate models could be overestimating the (heat-driven) moisture transport to the drier regions on land \citep{simpson23}. 

One way of assessing to what degree condensation-induced dynamics is taken into account in a given atmospheric context is by using the reversible setup (switching precipitation off by putting the condensate velocity equal to zero). No difference would indicate that condensation-induced dynamics is not captured in the model. Another way would be to use an artificial model where condensation is programmed as a temperature-dependent chemical reaction that does not change the amount of gas while releasing  the same amount of heat, and compare the output of this artificial model with the output of more realistic models.

The bottomline is that finding the right combination of model parameters that would capture real atmospheric dynamics is not possible without recognizing that besides heat there is a distinct driver of atmospheric motions, the condensation mass sink.  What determines the change between regimes when precipitation reduces or increases surface pressure (Fig.~\ref{figadj})? Is invigoration of convection by aerosols \citep{abbott2021} an example of such a switch?
We believe that focused theoretical and empirical studies of condensation-induced atmospheric dynamics are indispensable and urgent.

\section*{\large Acknowledgments}
{\large We thank Dr. Ruben Molina for useful discussions. Work of A.M. Makarieva is partially funded by the Federal Ministry of Education and Research (BMBF) and the Free State of Bavaria under the Excellence Strategy of the Federal Government and the L\"ander, as well as by the Technical University of Munich -- Institute for Advanced Study.}

\section*{\large Datastatement}
The raw data utilised in this study are available at Zenodo \url{https://doi.org/10.5281/zenodo.10577109}. These data were derived from the following resources available in the public domain:
\url{https://disc.gsfc.nasa.gov/datasets/TRMM\_3B42\_7/summary} and \url{https://rammb2.cira.colostate.edu/research/tropical-cyclones/tc\_extended\_best\_track\_dataset/}.

\appendix
\setcounter{section}{0}%
\section{\large Mass balance in the atmospheric air column}
\label{A0}

\setcounter{equation}{0}%
\renewcommand{\theequation}{A\arabic{equation}}%

We consider the storm as an axisymmetric vortex. The air mass $M$(kg m$^{-2}$) per unit surface area is given by the integral over the volume $V= H S$ of a cylinder with a height $H$ and a base of area $S$:
\begin{equation}\label{mas}
 M \equiv \frac{1}{S} \int_{V} \rho dV ,
\end{equation}
where $\rho=\rho_{d}+\rho_{\upsilon}$ is the moist air density (dry air and water vapor).

The convergence  $C$ of air in the considered region is determined as follows:
\begin{equation}\label{conv}
C\equiv  - \frac{1}{S} \int_{V} \mathrm{div} (\rho \mathbf{u}) dV = - \frac{1}{S} \oint_{\sigma}  \rho u_n d\sigma \equiv F_{\rm in} - F_{\rm out} ,
\end{equation}
where  $\mathbf{u}$ is the vector of air velocity. Its component  $u_n = (\mathbf{u}\cdot \mathbf{n})$ is  directed along the outward unit normal $\mathbf{n}$ to the closed surface $\sigma$, which encloses the volume $V$. Flux $F_{\rm in}$ corresponds to the case when the air flows into the region ($u_n <0$).  Flux $F_{\rm out}$ corresponds to the case when the air flows out of the region ($u_n  > 0$).  Since $\rho \simeq 0$  at the height $z = H$ and $u_n = 0$ at the Earth{\textquoteright}s surface, atmospheric air convergence $C$ describes the net flux of air across the lateral surface of the
cylinder.  

The difference between evaporation $E$ and precipitation $P$ is
\begin{equation}\label{dpe}
E -P   \equiv \frac{1}{S} \int \limits_{V} \dot{\rho}_{\upsilon}  dV  \simeq - P ,
\end{equation}
where $\dot{\rho}_{\upsilon}$ is the mass source/sink of water vapor. The  contribution of evaporation ($\dot{\rho}_{\upsilon} > 0$) is assumed to be negligibly  small compared to that of condensation ($\dot{\rho}_{\upsilon} <0$), i.e., $E \ll P$.

The surface pressure tendency is determined by the rate of change of air mass \eqref{mas}:
\begin{equation}\label{prt}
\frac{1}{g} \frac{\partial p}{\partial t} = \frac{\partial M}{\partial t} = C - P = F_{\rm in} - F_{\rm out} - P,  
\end{equation}
where $g$ is the acceleration of gravity. In eq.~\eqref{prt}, the second equality takes into account the mass balance in the atmospheric column.

We assume that flux $F_{\rm in}$ flows into the volume through a surface $\sigma_{\rm in}$ of area $S_{\rm in}= 2 \pi r h_{\rm in}$, while flux $F_{\rm out}$ flows out of the volume through a surface $\sigma_{\rm out}$ of area $S_{\rm out}= 2 \pi r h_{\rm out}$. Here  $h_{\rm in}$ and $h_{\rm out}$  denote the depth of the corresponding layers and  $r$ is the radius of cylinder. Surfaces $\sigma_{\rm in}$  and $\sigma_{\rm out}$ can be spatially separated. The average values of the air fluxes can be written as follows
\begin{equation}\label{flx}
F_{\rm in} = \rho u_{\rm in} \frac{S_{\rm in}}{S} = 2 \rho u_{\rm in} \frac{h_{\rm in}}{r} , \quad F_{\rm out} = \rho u_{\rm out} \frac{S_{\rm out}}{S} = 2 \rho u_{\rm out} \frac{h_{\rm out}}{r} , 
\end{equation}
where $S = \pi r^2$ and average velocities $u_{\rm in} = - u_n >0$ and $u_{\rm out} = u_n >0$.


\begin{thebibliography}{79}
\providecommand{\natexlab}[1]{#1}
\providecommand{\url}[1]{{\tt #1}}
\providecommand{\urlprefix}{URL }
\expandafter\ifx\csname urlstyle\endcsname\relax
  \providecommand{\doi}[1]{https://doi.org/\discretionary{}{}{}#1}\else
  \providecommand{\doi}{https://doi.org/\discretionary{}{}{}\begingroup
  \urlstyle{rm}\Url}\fi

\bibitem[{Abbott and Cronin(2021)}]{abbott2021}
Abbott, T.~H. and Cronin, T.~W.: Aerosol invigoration of atmospheric convection
  through increases in humidity, Science, 371, 83--85,
  \doi{10.1126/science.abc5181}, 2021.

\bibitem[{Aberson et~al.(2006)Aberson, Montgomery, Bell, and Black}]{aberson06}
Aberson, S.~D., Montgomery, M.~T., Bell, M., and Black, M.: Hurricane {Isabel}
  (2003): {New} insights into the physics of intense storms. {Part II:}
  {Extreme} localized wind, Bull. Amer. Meteor. Soc., 87, 1349--1354,
  \doi{10.1175/BAMS-87-10-1349}, 2006.

\bibitem[{An(2011)}]{an11}
An, S.-I.: Atmospheric responses of {Gill}-type and {Lindzen-Nigam} models to
  global warming, J. Climate, 24, 6165--6173, \doi{10.1175/2011JCLI3971.1},
  2011.

\bibitem[{Boos and Storelvmo(2016{\natexlab{a}})}]{Boos2016a}
Boos, W.~R. and Storelvmo, T.: Near-linear response of mean monsoon strength to
  a broad range of radiative forcings, Proc. Natl. Acad. Sci. USA, 113,
  1510--1515, \doi{10.1073/pnas.1517143113}, 2016{\natexlab{a}}.

\bibitem[{Boos and Storelvmo(2016{\natexlab{b}})}]{Boos2016b}
Boos, W.~R. and Storelvmo, T.: Reply to {Levermann et al.: Linear} scaling for
  monsoons based on well-verified balance between adiabatic cooling and latent
  heat release, Proc. Natl. Acad. Sci. USA, 113, E2350--E2351,
  \doi{10.1073/pnas.1603626113}, 2016{\natexlab{b}}.

\bibitem[{Bryan and Fritsch(2002)}]{br02}
Bryan, G.~H. and Fritsch, J.~M.: A benchmark simulation for moist
  nonhydrostatic numerical models, Mon. Wea. Rev., 130, 2917--2928,
  \doi{10.1175/1520-0493(2002)130<2917:ABSFMN>2.0.CO;2}, 2002.

\bibitem[{Bryan and Rotunno(2009)}]{bryan09a}
Bryan, G.~H. and Rotunno, R.: The maximum intensity of tropical cyclones in
  axisymmetric numerical model simulations, Mon. Wea. Rev., 137, 1770--1789,
  \doi{10.1175/2008MWR2709.1}, 2009.

\bibitem[{Chagnon and Bannon(2005)}]{chagnon05}
Chagnon, J.~M. and Bannon, P.~R.: Adjustment to injections of mass, momentum,
  and heat in a compressible atmosphere, J. Atmos. Sci., 62, 2749--2769,
  \doi{10.1175/JAS3503.1}, 2005.

\bibitem[{Chen and Zhang(2013)}]{chen13}
Chen, H. and Zhang, D.-L.: On the rapid intensification of {Hurricane Wilma
  (2005). Part II: Convective} bursts and the upper-level warm core, J. Atmos.
  Sci., 70, 146--162, \doi{10.1175/JAS-D-12-062.1}, 2013.

\bibitem[{Daloz et~al.(2015)Daloz, Camargo, Kossin, Emanuel, Horn, Jonas, Kim,
  {LaRow}, Lim, Patricola, Roberts, Scoccimarro, Shaevitz, Vidale, Wang,
  Wehner, and Zhao}]{daloz15}
Daloz, A.~S., Camargo, S.~J., Kossin, J.~P., Emanuel, K., Horn, M., Jonas,
  J.~A., Kim, D., {LaRow}, T., Lim, Y.-K., Patricola, C.~M., Roberts, M.,
  Scoccimarro, E., Shaevitz, D., Vidale, P.~L., Wang, H., Wehner, M., and Zhao,
  M.: Cluster analysis of downscaled and explicitly simulated {North Atlantic}
  tropical cyclone tracks, J. Climate, 28, 1333--1361,
  \doi{10.1175/JCLI-D-13-00646.1}, 2015.

\bibitem[{Davison and Haynes(2022)}]{davison22}
Davison, M. and Haynes, P.: Excitable {Madden-Julian Oscillation} like
  behaviour of a simple model of equatorial moist dynamics results from a time
  step that is too large, Quart. J. Roy. Meteor. Soc., 148, 770--777,
  \doi{10.1002/qj.4229}, 2022.

\bibitem[{Demuth et~al.(2006)Demuth, {DeMaria}, and Knaff}]{demuth06}
Demuth, J.~L., {DeMaria}, M., and Knaff, J.~A.: Improvement of advanced
  microwave sounding unit tropical cyclone intensity and size estimation
  algorithms, J. Appl. Meteor. Climatol., 45, 1573--1581,
  \doi{10.1175/JAM2429.1}, 2006.

\bibitem[{Emanuel(1988)}]{em88}
Emanuel, K.~A.: The maximum intensity of hurricanes, J. Atmos. Sci., 45,
  1143--1155, \doi{10.1175/1520-0469(1988)045<1143:TMIOH>2.0.CO;2}, 1988.

\bibitem[{Emanuel(1991)}]{emanuel91}
Emanuel, K.~A.: The theory of hurricanes, Annu. Rev. Fluid Mech., 23, 179--196,
  \doi{10.1146/annurev.fl.23.010191.001143}, 1991.

\bibitem[{Emanuel(1999)}]{emanuel99}
Emanuel, K.~A.: The power of a hurricane: An example of reckless driving on the
  information superhighway, Weather, 54, 107--108,
  \doi{10.1002/j.1477-8696.1999.tb06435.x}, 1999.

\bibitem[{Emanuel et~al.(1994)Emanuel, Neelin, and Bretherton}]{emetal94}
Emanuel, K.~A., Neelin, J.~D., and Bretherton, C.~S.: On large-scale
  circulations in convecting atmospheres, Quart. J. Roy. Meteor. Soc., 120,
  1111--1143, \doi{10.1002/qj.49712051902}, 1994.

\bibitem[{Emanuel et~al.(1997)Emanuel, Neelin, and Bretherton}]{emanuel97}
Emanuel, K.~A., Neelin, J.~D., and Bretherton, C.~S.: Reply to comments by
  {Bjorn Stevens, David A. Randall, Xin Lin and Michael T. Montgomery} on
  {\textquoteleft}{On} large-scale circulations in convecting
  atmospheres{\textquoteright} {(July B, 1994, \textbf{120}, 1111--1143)},
  Quart. J. Roy. Meteor. Soc., 123, 1779--1782, \doi{10.1002/qj.49712354217},
  1997.

\bibitem[{Frank(1977)}]{frank77}
Frank, W.~M.: The structure and energetics of the tropical cyclone {I. Storm}
  structure, Mon. Wea. Rev., 105, 1119--1135,
  \doi{10.1175/1520-0493(1977)105<1119:TSAEOT>2.0.CO;2}, 1977.

\bibitem[{Goody(2003)}]{goody03}
Goody, R.: On the mechanical efficiency of deep, tropical convection, J. Atmos.
  Sci., 60, 2827--2832, \doi{10.1175/1520-0469(2003)060<2827:OTMEOD>2.0.CO;2},
  2003.

\bibitem[{Gorshkov et~al.(2012)Gorshkov, Makarieva, and Nefiodov}]{jetp12}
Gorshkov, V.~G., Makarieva, A.~M., and Nefiodov, A.~V.: Condensation of water
  vapor in the gravitational field, J. Exp. Theor. Phys., 115, 723--728,
  \doi{10.1134/S106377611209004X}, 2012.

\bibitem[{He et~al.(2018)He, Chan, and Li}]{he18}
He, Y.~C., Chan, P.~W., and Li, Q.~S.: Observational study on thermodynamic and
  kinematic structures of {Typhoon Vicente} (2012) at landfall, J. Wind Eng.
  Ind. Aerodyn., 172, 280--297, \doi{10.1016/j.jweia.2017.11.008}, 2018.

\bibitem[{He et~al.(2019)He, Li, Chan, Fu, Wu, and Li}]{he19}
He, Y.~C., Li, Y.~Z., Chan, P.~W., Fu, J.~Y., Wu, J.~R., and Li, Q.~S.: A
  height-resolving model of tropical cyclone pressure field, J. Wind Eng. Ind.
  Aerodyn., 186, 84--93, \doi{10.1016/j.jweia.2018.12.020}, 2019.

\bibitem[{Holland(1980)}]{holland80}
Holland, G.~J.: An analytic model of the wind and pressure profiles in
  hurricanes, Mon. Wea. Rev., 108, 1212--1218,
  \doi{10.1175/1520-0493(1980)108<1212:AAMOTW>2.0.CO;2}, 1980.

\bibitem[{Holland et~al.(2010)Holland, Belanger, and Fritz}]{holland10}
Holland, G.~J., Belanger, J.~I., and Fritz, A.: A revised model for radial
  profiles of hurricane winds, Mon. Wea. Rev., 138, 4393--4401,
  \doi{10.1175/2010MWR3317.1}, 2010.

\bibitem[{Holliday and Thompson(1979)}]{holliday79}
Holliday, C.~R. and Thompson, A.~H.: Climatological characteristics of rapidly
  intensifying typhoons, Mon. Wea. Rev., 107, 1022--1034,
  \doi{10.1175/1520-0493(1979)107<1022:CCORIT>2.0.CO;2}, 1979.

\bibitem[{Huffman et~al.(2007)Huffman, Adler, Bolvin, Gu, Nelkin, Bowman, Hong,
  Stocker, and Wolff}]{huffman07}
Huffman, G.~J., Adler, R.~F., Bolvin, D.~T., Gu, G., Nelkin, E.~J., Bowman,
  K.~P., Hong, Y., Stocker, E.~F., and Wolff, D.~B.: The {TRMM Multisatellite
  Precipitation Analysis (TMPA):} Quasi-global, multiyear, combined-sensor
  precipitation estimates at fine scales, J. Hydrometeor., 8, 38--55,
  \doi{10.1175/JHM560.1}, 2007.

\bibitem[{Kaplan and DeMaria(2003)}]{kaplan03}
Kaplan, J. and DeMaria, M.: Large-scale characteristics of rapidly intensifying
  tropical cyclones in the {North Atlantic} basin, Wea. Forecasting, 18,
  1093--1108, \doi{10.1175/1520-0434(2003)018<1093:LCORIT>2.0.CO;2}, 2003.

\bibitem[{Kieu(2015)}]{kieu2015}
Kieu, C.: Hurricane maximum potential intensity equilibrium, Quart. J. Roy.
  Meteor. Soc., 141, 2471--2480, \doi{10.1002/qj.2556}, 2015.

\bibitem[{Kieu(2004)}]{kieu04}
Kieu, C.~Q.: An analytical theory for the early stage of the development of
  hurricanes: {Part I},
  \urlprefix\url{https://doi.org/10.48550/arXiv.physics/0407073}, eprint
  arXiv:physics/0407073v2[physics.ao-ph], 2004.

\bibitem[{Kieu(2006)}]{kieu06}
Kieu, C.~Q.: On the roles of the secondary circulation in the formation of
  hurricanes, \urlprefix\url{https://doi.org/10.48550/arXiv.physics/0610273},
  eprint arXiv:physics/0610273 [physics.ao-ph], 2006.

\bibitem[{Kieu and Zhang(2009)}]{kieu09}
Kieu, C.~Q. and Zhang, D.-L.: An analytical model for the rapid intensification
  of tropical cyclones, Quart. J. Roy. Meteor. Soc., 135, 1336--1349,
  \doi{10.1002/qj.433}, 2009.

\bibitem[{Kowch and Emanuel(2015)}]{kowch15}
Kowch, R. and Emanuel, K.: Are special processes at work in the rapid
  intensification of tropical cyclones?, Mon. Wea. Rev., 143, 878--882,
  \doi{10.1175/MWR-D-14-00360.1}, 2015.

\bibitem[{Kurihara(1975)}]{kurihara75}
Kurihara, Y.: Budget analysis of a tropical cyclone simulated in an
  axisymmetric numerical model, J. Atmos. Sci., 32, 25--59,
  \doi{10.1175/1520-0469(1975)032<0025:BAOATC>2.0.CO;2}, 1975.

\bibitem[{Lackmann and Yablonsky(2004)}]{la04}
Lackmann, G.~M. and Yablonsky, R.~M.: The importance of the precipitation mass
  sink in tropical cyclones and other heavily precipitating systems, J. Atmos.
  Sci., 61, 1674--1692, \doi{10.1175/1520-0469(2004)061<1674:TIOTPM>2.0.CO;2},
  2004.

\bibitem[{Lee et~al.(2016)Lee, Tippett, Sobel, and Camargo}]{lee16}
Lee, C.-Y., Tippett, M.~K., Sobel, A.~H., and Camargo, S.~J.: Rapid
  intensification and the bimodal distribution of tropical cyclone intensity,
  Nat. Commun., 7, \doi{10.1038/ncomms10625}, 2016.

\bibitem[{Levermann et~al.(2009)Levermann, Schewe, Petoukhov, and
  Held}]{Levermann2009}
Levermann, A., Schewe, J., Petoukhov, V., and Held, H.: Basic mechanism for
  abrupt monsoon transitions, Proc. Natl. Acad. Sci. USA, 106,
  20\,572--20\,577, \doi{10.1073/pnas.0901414106}, 2009.

\bibitem[{Levermann et~al.(2016)Levermann, Petoukhov, Schewe, and
  Schellnhuber}]{Levermann2016}
Levermann, A., Petoukhov, V., Schewe, J., and Schellnhuber, H.~J.: Abrupt
  monsoon transitions as seen in paleorecords can be explained by
  moisture-advection feedback, Proc. Natl. Acad. Sci. USA, 113, E2348--E2349,
  \doi{10.1073/pnas.1603130113}, 2016.

\bibitem[{Li et~al.(2021)Li, Wang, Lin, and Wang}]{li21}
Li, Y., Wang, Y., Lin, Y., and Wang, X.: Why does rapid contraction of the
  radius of maximum wind precede rapid intensification in tropical cyclones?,
  J. Atmos. Sci., 78, 3441--3453, \doi{10.1175/JAS-D-21-0129.1}, 2021.

\bibitem[{Lindzen and Nigam(1987)}]{lindzen87}
Lindzen, R.~S. and Nigam, S.: On the role of sea surface temperature gradients
  in forcing low-level winds and convergence in the tropics, J. Atmos. Sci.,
  44, 2418--2436, \doi{10.1175/1520-0469(1987)044<2418:OTROSS>2.0.CO;2}, 1987.

\bibitem[{Lonfat et~al.(2004)Lonfat, {Marks Jr.}, and Chen}]{lonfat04}
Lonfat, M., {Marks Jr.}, F.~D., and Chen, S.~S.: Precipitation distribution in
  tropical cyclones using the {Tropical Rainfall Measuring Mission (TRMM)}
  microwave imager: {A} global perspective, Mon. Wea. Rev., 132, 1645--1660,
  \doi{10.1175/1520-0493(2004)132<1645:PDITCU>2.0.CO;2}, 2004.

\bibitem[{Makarieva and Gorshkov(2007)}]{hess07}
Makarieva, A.~M. and Gorshkov, V.~G.: Biotic pump of atmospheric moisture as
  driver of the hydrological cycle on land, Hydrol. Earth Syst. Sci., 11,
  1013--1033, \doi{10.5194/hess-11-1013-2007}, 2007.

\bibitem[{Makarieva and Gorshkov(2009)}]{pla09b}
Makarieva, A.~M. and Gorshkov, V.~G.: {Condensation-induced} kinematics and
  dynamics of cyclones, hurricanes and tornadoes, Phys. Lett. A, 373,
  4201--4205, \doi{10.1016/j.physleta.2009.09.023}, 2009.

\bibitem[{Makarieva and Gorshkov(2011)}]{pla11}
Makarieva, A.~M. and Gorshkov, V.~G.: Radial profiles of velocity and pressure
  for {condensation-induced} hurricanes, Phys. Lett. A, 375, 1053--1058,
  \doi{10.1016/j.physleta.2011.01.005}, 2011.

\bibitem[{Makarieva et~al.(2011)Makarieva, Gorshkov, and Nefiodov}]{tor11}
Makarieva, A.~M., Gorshkov, V.~G., and Nefiodov, A.~V.: Condensational theory
  of stationary tornadoes, Phys. Lett. A, 375, 2259--2261,
  \doi{10.1016/j.physleta.2011.04.023}, 2011.

\bibitem[{Makarieva et~al.(2013{\natexlab{a}})Makarieva, Gorshkov, Nefiodov,
  Sheil, Nobre, Bunyard, and Li}]{jas13}
Makarieva, A.~M., Gorshkov, V.~G., Nefiodov, A.~V., Sheil, D., Nobre, A.~D.,
  Bunyard, P., and Li, B.-L.: The key physical parameters governing frictional
  dissipation in a precipitating atmosphere, J. Atmos. Sci., 70, 2916--2929,
  \doi{10.1175/JAS-D-12-0231.1}, 2013{\natexlab{a}}.

\bibitem[{Makarieva et~al.(2013{\natexlab{b}})Makarieva, Gorshkov, Sheil,
  Nobre, and Li}]{acp13}
Makarieva, A.~M., Gorshkov, V.~G., Sheil, D., Nobre, A.~D., and Li, B.-L.:
  Where do winds come from? {A} new theory on how water vapor condensation
  influences atmospheric pressure and dynamics, Atmos. Chem. Phys., 13,
  1039--1056, \doi{10.5194/acp-13-1039-2013}, 2013{\natexlab{b}}.

\bibitem[{Makarieva et~al.(2015)Makarieva, Gorshkov, and Nefiodov}]{pla15}
Makarieva, A.~M., Gorshkov, V.~G., and Nefiodov, A.~V.: Empirical evidence for
  the condensational theory of hurricanes, Phys. Lett. A, 379, 2396--2398,
  \doi{10.1016/j.physleta.2015.07.042}, 2015.

\bibitem[{Makarieva et~al.(2017{\natexlab{a}})Makarieva, Gorshkov, Nefiodov,
  Chikunov, Sheil, Nobre, and Li}]{ar17}
Makarieva, A.~M., Gorshkov, V.~G., Nefiodov, A.~V., Chikunov, A.~V., Sheil, D.,
  Nobre, A.~D., and Li, B.-L.: Fuel for cyclones: {The} water vapor budget of a
  hurricane as dependent on its movement, Atmos. Res., 193, 216--230,
  \doi{10.1016/j.atmosres.2017.04.006}, 2017{\natexlab{a}}.

\bibitem[{Makarieva et~al.(2017{\natexlab{b}})Makarieva, Gorshkov, Nefiodov,
  Sheil, Nobre, and Li}]{arxiv17}
Makarieva, A.~M., Gorshkov, V.~G., Nefiodov, A.~V., Sheil, D., Nobre, A.~D.,
  and Li, B.-L.: Quantifying the global atmospheric power budget,
  \urlprefix\url{https://arxiv.org/abs/1603.03706}, eprint arXiv: 1603.03706v4
  [physics.ao-ph], 2017{\natexlab{b}}.

\bibitem[{Makarieva et~al.(2023)Makarieva, Gorshkov, Nefiodov, Chikunov, Sheil,
  Nobre, Nobre, Plunien, and Molina}]{mpi4-jas}
Makarieva, A.~M., Gorshkov, V.~G., Nefiodov, A.~V., Chikunov, A.~V., Sheil, D.,
  Nobre, A.~D., Nobre, P., Plunien, G., and Molina, R.~D.: Water lifting and
  outflow gain of kinetic energy in tropical cyclones, J. Atmos. Sci., 80,
  1905--1921, \doi{10.1175/JAS-D-21-0172.1}, 2023.

\bibitem[{Masunaga and Mapes(2020)}]{masunaga2020}
Masunaga, H. and Mapes, B.~E.: A mechanism for the maintenance of sharp
  tropical margins, J. Atmos. Sci., 77, 1181--1197,
  \doi{10.1175/JAS-D-19-0154.1}, 2020.

\bibitem[{Miller(1958)}]{miller58}
Miller, B.~I.: The three-dimensional wind structure around a tropical cyclone,
  Nat. Hurricane Res. Proj. Rep. 15, U.S. Dep. of Commer., Washington, D.C.,
  1958.

\bibitem[{Montgomery et~al.(2006)Montgomery, Bell, Aberson, and
  Black}]{montgomery06}
Montgomery, M.~T., Bell, M.~M., Aberson, S.~D., and Black, M.~L.: Hurricane
  {Isabel} (2003): {New} insights into the physics of intense storms. {Part I:}
  {Mean} vortex structure and maximum intensity estimates, Bull. Amer. Meteor.
  Soc., 87, 1335--1348, \doi{10.1175/BAMS-87-10-1335}, 2006.

\bibitem[{Montgomery et~al.(2009)Montgomery, Sang, Smith, and
  Persing}]{montgomery09}
Montgomery, M.~T., Sang, N.~V., Smith, R.~K., and Persing, J.: Do tropical
  cyclones intensify by {WISHE?}, Quart. J. Roy. Meteor. Soc., 135, 1697--1714,
  \doi{10.1002/qj.459}, 2009.

\bibitem[{Neelin(1989)}]{neelin89}
Neelin, J.~D.: On the interpretation of the {Gill} model, J. Atmos. Sci., 46,
  2466--2468, \doi{10.1175/1520-0469(1989)046<2466:OTIOTG>2.0.CO;2}, 1989.

\bibitem[{Qiu et~al.(1993)Qiu, Bao, and Xu}]{qiu93}
Qiu, C.-J., Bao, J.-W., and Xu, Q.: Is the mass sink due to precipitation
  negligible?, Mon. Wea. Rev., 121, 853--857,
  \doi{10.1175/1520-0493(1993)121<0853:ITMSDT>2.0.CO;2}, 1993.

\bibitem[{Rodgers and Adler(1981)}]{rodgers81}
Rodgers, E.~B. and Adler, R.~F.: Tropical cyclone rainfall characteristics as
  determined from a satellite passive microwave radiometer, J. Atmos. Sci.,
  109, 206--221, \doi{10.1175/1520-0493(1981)109<0506:TCRCAD>2.0.CO;2}, 1981.

\bibitem[{Rotunno and Emanuel(1987)}]{ro87}
Rotunno, R. and Emanuel, K.~A.: An air-sea interaction theory for tropical
  cyclones. {Part II}: {Evolutionary} study using a nonhydrostatic axisymmetric
  numerical model, J. Atmos. Sci., 44, 542--561,
  \doi{10.1175/1520-0469(1987)044<0542:AAITFT>2.0.CO;2}, 1987.

\bibitem[{Rousseau-Rizzi et~al.(2021)Rousseau-Rizzi, Rotunno, and
  Bryan}]{rizzi21}
Rousseau-Rizzi, R., Rotunno, R., and Bryan, G.: A thermodynamic perspective on
  steady-state tropical cyclones, J. Atmos. Sci., 78, 583--593,
  \doi{10.1175/JAS-D-20-0140.1}, 2021.

\bibitem[{Schubert and Hack(1982)}]{schubert82}
Schubert, W.~H. and Hack, J.~J.: Inertial stability and tropical cyclone
  development, J. Atmos. Sci., 39, 1687--1697,
  \doi{10.1175/1520-0469(1982)039<1687:ISATCD>2.0.CO;2}, 1982.

\bibitem[{Simpson et~al.(2023)Simpson, {McKinnon}, Kennedy, Lawrence, Lehner,
  and Seager}]{simpson23}
Simpson, I.~R., {McKinnon}, K.~A., Kennedy, D., Lawrence, D.~M., Lehner, F.,
  and Seager, R.: Observed humidity trends in dry regions contradict climate
  models, Proc. Natl. Acad. Sci. USA, 121, e2302480\,120,
  \doi{10.1073/pnas.2302480120}, 2023.

\bibitem[{Smith and Montgomery(2016{\natexlab{a}})}]{smith16}
Smith, R.~K. and Montgomery, M.~T.: Understanding hurricanes, Weather, 71,
  219--223, \doi{10.1002/wea.2776}, 2016{\natexlab{a}}.

\bibitem[{Smith and Montgomery(2016{\natexlab{b}})}]{smith16b}
Smith, R.~K. and Montgomery, M.~T.: The efficiency of diabatic heating and
  tropical cyclone intensification, Quart. J. Roy. Meteor. Soc., 142,
  2081--2086, \doi{10.1002/qj.2804}, 2016{\natexlab{b}}.

\bibitem[{Smith et~al.(2018)Smith, Montgomery, and Kilroy}]{smith18}
Smith, R.~K., Montgomery, M.~T., and Kilroy, G.: The generation of kinetic
  energy in tropical cyclones revisited, Quart. J. Roy. Meteor. Soc., 144,
  2481--2490, \doi{10.1002/qj.3332}, 2018.

\bibitem[{Sparks and Toumi(2022{\natexlab{a}})}]{sparks22a}
Sparks, N. and Toumi, R.: A physical model of tropical cyclone central pressure
  filling at landfall, J. Atmos. Sci., 79, 2585--2599,
  \doi{10.1175/JAS-D-21-0196.1}, 2022{\natexlab{a}}.

\bibitem[{Sparks and Toumi(2022{\natexlab{b}})}]{sparks22b}
Sparks, N. and Toumi, R.: The dependence of tropical cyclone pressure tendency
  on size, Geophys. Res. Lett., 49, e2022GL098\,926,
  \doi{10.1029/2022GL098926}, 2022{\natexlab{b}}.

\bibitem[{Spengler et~al.(2011)Spengler, Egger, and Garner}]{spengler11}
Spengler, T., Egger, J., and Garner, S.~T.: How does rain affect surface
  pressure in a one-dimensional framework?, J. Atmos. Sci., 68, 347--360,
  \doi{10.1175/2010JAS3582.1}, 2011.

\bibitem[{Stevens et~al.(1997)Stevens, Randall, Lin, and
  Montgomery}]{stevens97}
Stevens, B., Randall, D.~A., Lin, X., and Montgomery, M.~T.: Comments on
  {\textquoteleft}{On} large-scale circulations in convecting
  atmospheres{\textquoteright} by {Kerry A. Emanuel, J. David Neelin and
  Christopher S. Bretherton} {(July B, 1994, \textbf{120}, 1111--1143)}, Quart.
  J. Roy. Meteor. Soc., 123, 1771--1778, \doi{10.1002/qj.49712354216}, 1997.

\bibitem[{Tu et~al.(2021)Tu, Xu, Chan, Huang, Xu, and Chiu}]{tu21}
Tu, S., Xu, J., Chan, J. C.~L., Huang, K., Xu, F., and Chiu, L.~S.: Recent
  global decrease in the inner-core rain rate of tropical cyclones, Nat.
  Commun., 12, \doi{10.1038/s41467-021-22304-y}, 2021.

\bibitem[{Vallis(2022)}]{vallis22}
Vallis, G.~K.: Reply to the comment by {Davison and Haynes: Madden-Julian
  Oscillation} like behaviour does persist at small time steps, Quart. J. Roy.
  Meteor. Soc., 148, 1127--1130, \doi{10.1002/qj.4250}, 2022.

\bibitem[{Vallis and Penn(2020)}]{vallis20}
Vallis, G.~K. and Penn, J.: Convective organization and eastward propagating
  equatorial disturbances in a simple excitable system, Quart. J. Roy. Meteor.
  Soc., 146, 2297--2314, \doi{10.1002/qj.3792}, 2020.

\bibitem[{van~den Dool and Saha(1993)}]{dool93}
van~den Dool, H.~M. and Saha, S.: Seasonal redistribution and conservation of
  atmospheric mass in a general circulation model, J. Climate, 6, 22--30,
  \doi{10.1175/1520-0442(1993)006<0022:SRACOA>2.0.CO;2}, 1993.

\bibitem[{Vickery and Wadhera(2008)}]{vickery08}
Vickery, P.~J. and Wadhera, D.: Statistical models of {Holland} pressure
  profile parameter and radius to maximum winds of hurricanes from flight-level
  pressure and {H*Wind} data, J. Appl. Meteor. Climatol., 47, 2497--2517,
  \doi{10.1175/2008JAMC1837.1}, 2008.

\bibitem[{Wang and Lin(2020)}]{wang20}
Wang, D. and Lin, Y.: Size and structure of dry and moist reversible tropical
  cyclones, J. Atmos. Sci., 77, 2091--2114, \doi{10.1175/JAS-D-19-0229.1},
  2020.

\bibitem[{Wang and Lin(2021)}]{wang21}
Wang, D. and Lin, Y.: Potential role of irreversible moist processes in
  modulating tropical cyclone surface wind structure, J. Atmos. Sci., 78, 709
  -- 725, \doi{10.1175/JAS-D-20-0192.1}, 2021.

\bibitem[{Wang et~al.(2024)Wang, Jiang, and Guzman}]{wang24}
Wang, X., Jiang, H., and Guzman, O.: Relating tropical cyclone intensification
  rate to precipitation and convective features in the inner core, Wea.
  Forecasting, \doi{10.1175/WAF-D-23-0155.1}, in press, 2024.

\bibitem[{Willoughby(1990)}]{willoughby90}
Willoughby, H.~E.: Temporal changes of the primary circulation in tropical
  cyclones, J. Atmos. Sci., 47, 242--264,
  \doi{10.1175/1520-0469(1990)047<0242:TCOTPC>2.0.CO;2}, 1990.

\bibitem[{Wu and Ruan(2021)}]{wu21}
Wu, Q. and Ruan, Z.: Rapid contraction of the radius of maximum tangential wind
  and rapid intensification of a tropical cyclone, J. Geophys. Res. Atmos.,
  126, e2020JD033\,681, \doi{10.1029/2020JD033681}, 2021.

\bibitem[{Yoon(2023)}]{yoon23}
Yoon, A.: The impact of {Amazon} deforestation on rain system using a
  storm-resolving global climate model,
  \urlprefix\url{https://doi.org/10.5194/egusphere-egu23-1304}, {EGU General
  Assembly 2023, Vienna, Austria, 24--28 Apr 2023, EGU23-1304}, 2023.

\end{thebibliography}

}

\end{document}